\documentclass[nofootinbib,aps,prl,preprint]{revtex4-1}
\usepackage{amsmath}
\usepackage{bbold}
\usepackage{graphicx}
\usepackage[hidelinks]{hyperref}
\usepackage{bookmark}% faster updated bookmarks
\usepackage{lineno}
\newcommand{\etal}{\textit{et al}.}

\newcommand{\parallelsum}{\mathbin{\!/\mkern-5mu/\!}}
\newcommand{\beginsupplement}{%
        \setcounter{table}{0}
        \renewcommand{\thetable}{S\arabic{table}}%
        \setcounter{figure}{0}
        \renewcommand{\thefigure}{S\arabic{figure}}%
     }

\begin{document}
\title{High-field depinned phase and planar Hall effect in skyrmion-host Gd$_2$PdSi$_3$}
\author{Max Hirschberger$^{1,2}$}\email{hirschberger@ap.t.u-tokyo.ac.jp}
\author{Taro Nakajima$^{2,3}$}\altaffiliation{Current address: Institute for Solid State Physics, The University of Tokyo, Kashiwa, Chiba 2778561, Japan}
\author{Markus Kriener$^{2}$}
\author{Takashi Kurumaji$^{2}$}\altaffiliation{Current address: Department of Physics, Massachusetts Institute of Technology, Cambridge, Massachusetts 02139, USA}
\author{Leonie Spitz$^{2,4}$}\altaffiliation{Current address: Physik-Department E51, Technische Universit{\"a}t M{\"u}nchen, D-85748 Garching, Germany}
\author{Shang Gao$^{2}$}
\author{Akiko Kikkawa$^{2}$}
\author{Yuichi Yamasaki$^{5,6}$}
\author{Hajime Sagayama$^{7}$}
\author{Hironori Nakao$^{7}$}
\author{Seiko Ohira-Kawamura$^{8}$}
\author{Yasujiro Taguchi$^{2}$}
\author{Taka-hisa Arima$^{3}$}
\author{Yoshinori Tokura$^{1,2,9}$}
\date{\today}
\affiliation{$^{1}$Department of Applied Physics and Quantum-Phase Electronics Center (QPEC), The University of Tokyo, Bunkyo-ku, Tokyo, 113-8656, Japan}
\affiliation{$^{2}$RIKEN Center for Emergent Matter Science (CEMS),Wako 351-0198, Japan}
\affiliation{$^{3}$Department of Advanced Materials Science, University of Tokyo, Kashiwa, Chiba 277-8561, Japan}
\affiliation{$^{4}$Physik-Department, Technical University of Munich, 85748 Garching, Germany}
\affiliation{$^{5}$Research and Services Division of Materials Data and Integrated System (MaDIS), National Institute for Materials Science (NIMS), Tsukuba 305-0047, Japan}
\affiliation{$^{6}$PRESTO, Japan Science and Technology Agency (JST), Kawaguchi 332-0012, Japan}
\affiliation{$^{7}$Institute of Materials Structure Science, High Energy Accelerator Research Organization, Tsukuba, Ibaraki 305-0801, Japan}
\affiliation{$^{8}$Materials and Life Science Division, J-PARC Center, Tokai, Ibaraki, 319-1195, Japan}
\affiliation{$^{9}$Tokyo College, The University of Tokyo, Bunkyo-ku 113-8656, Japan}

\begin{abstract}
For the skyrmion-hosting intermetallic Gd$_2$PdSi$_3$ with centrosymmetric hexagonal lattice and triangular net of rare earth sites, we report a thorough investigation of the magnetic phase diagram. Our work reveals a new magnetic phase with isotropic value of the critical field for all orientations, where the magnetic ordering vector $\mathbf{q}$ is depinned from its preferred directions in the basal plane. This is in contrast to the highly anisotropic behavior of the low field phases, such as the skyrmion lattice (SkL), which are easily destroyed by in-plane magnetic field. The bulk nature of the SkL and of other magnetic phases was evidenced by specific-heat measurements. Resistivity anisotropy, likely originating from partial gapping of the density of states along $\mathbf{q}$ in this RKKY magnet, is picked up via the planar Hall effect (PHE). The PHE confirms the single-$\mathbf{q}$ nature of the magnetic order when the field is in the hexagonal plane, and allows to detect the preferred directions of $\mathbf{q}$. For field aligned perpendicular to the basal plane, several scenarios for the depinned phase (DP), such as tilted conical order, are discussed on the basis of the data.
\end{abstract}

\maketitle
%trim: [from left edge, from bottom , from right edge, ..]
\begin{figure*}[htb]
  \begin{center}
	%trim: [from left edge, from bottom , from right edge, ..]
		\includegraphics[trim=0cm 0cm 0cm 0cm, width=1\linewidth]{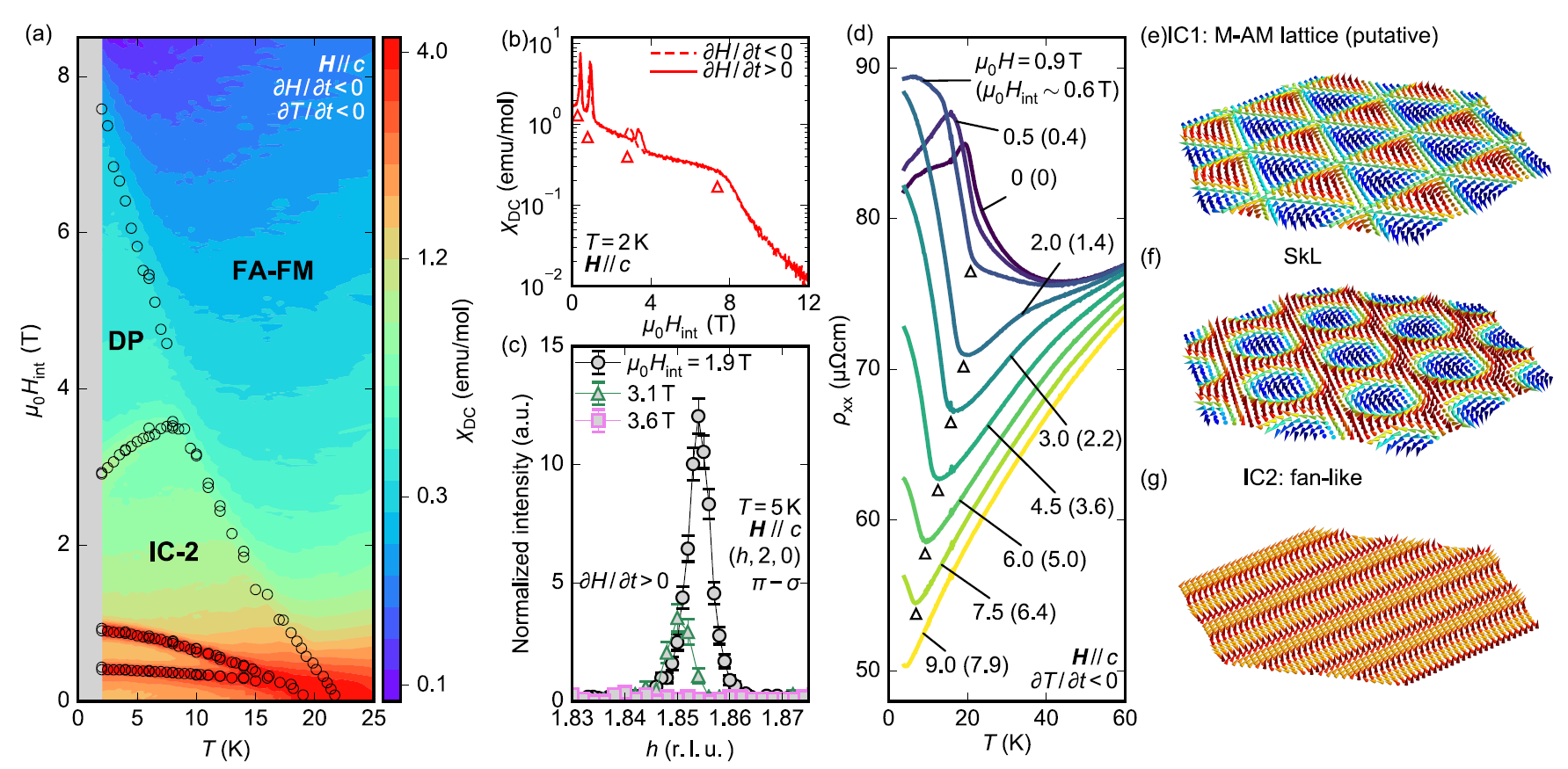}
    \caption[]{(color online). (a) DC susceptibility $\chi_\text{DC}=\partial M / \partial H$ on a logarithmic scale shown as a function of temperature $T$ and $\mathbf{H}\,\parallelsum\,c$. Phases IC-1 (meron-antimeron lattice M-AM), skyrmion lattice SkL, IC-2 (fan-like), DP (depinned high-field phase), and FA-FM (field-aligned ferromagnet) are indicated in these panels and illustrated in (e-g), where red (blue) colors correspond to moment up (down). Phase DP was not reported in previous work. (b) Example of raw data used to construct panels (a), with a sequence of phase transitions indicated by red triangles. Note the logarithmic $y$-axis. (c) $\pi$-$\sigma$ intensity observed in resonant elastic x-ray scattering (REXS) at the Gd-L$_2$ edge with HK0 scattering plane and horizontal geometry. While incommensurate magnetic order with $\mathbf{q} = (0.14, 0, 0)$ could be detected in phase IC-2, no signal appears at this position in phase DP (magenta curve). (d) Resistivity $\rho_{xx}$ as a function of $T$ for different values of $\mu_0 H$. The corresponding values for $H_\text{int}$ are calculated at $T = 2\,$K. The data show sharp kinks at the onset of long-range order (black open triangles).}
    \label{fig:fig1}
  \end{center}
\end{figure*}
\begin{figure}[b!]
  \begin{center}
	%trim: [from left edge, from bottom , from right edge, ..]
%    \includegraphics[clip, trim=9.0cm 2.5cm 9.0cm 0cm, width=1.0\linewidth]{FIG1.pdf}
		\includegraphics[trim=0.7cm 0.1cm 0.7cm 0.1cm, width=0.7\linewidth]{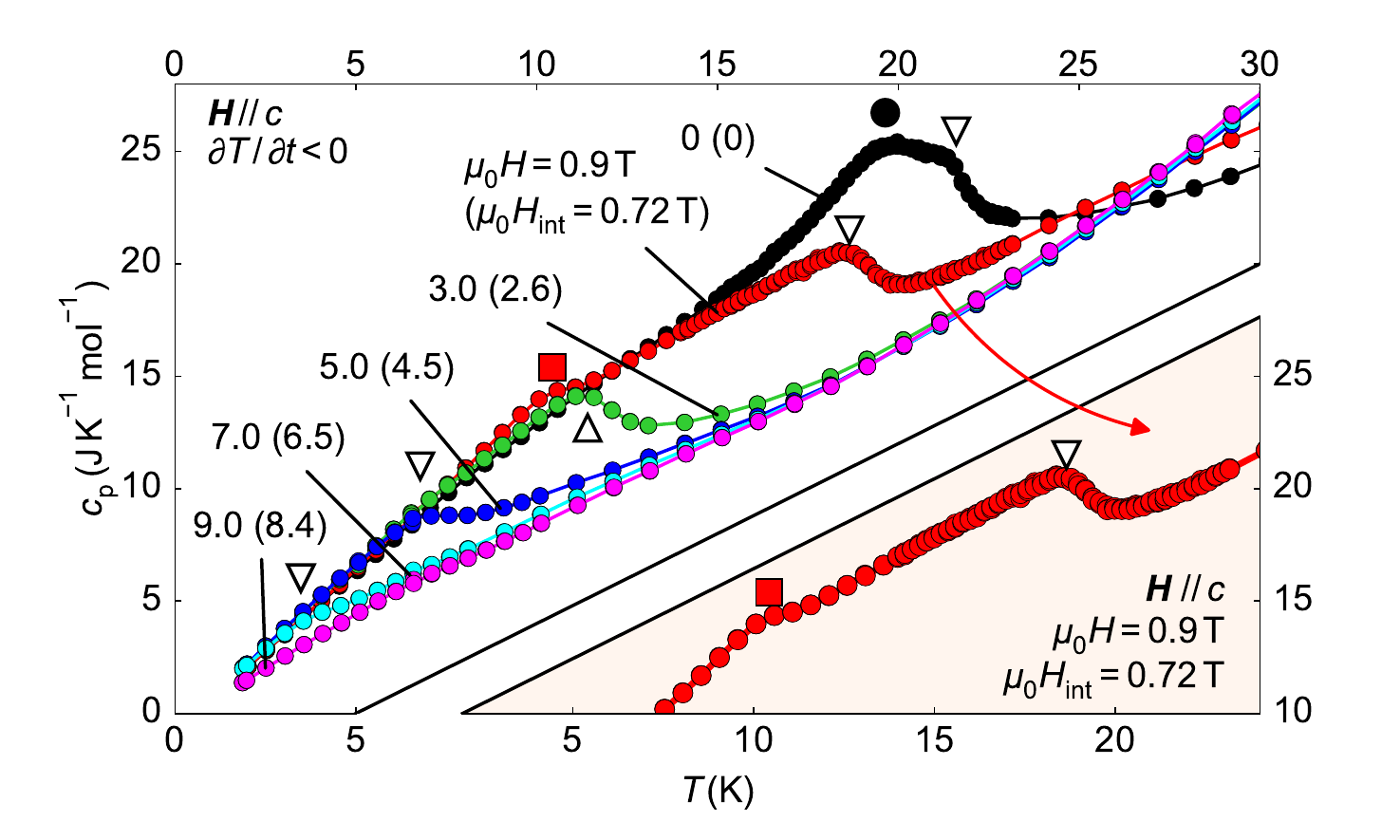}
    \caption[]{(color online). Specific heat of Gd$_2$PdSi$_3$. The boundaries of IC-2 and DP were marked by black open triangles, the upper boundary of IC-1 by a black disk, and the onset of phase SkL by a red square. Inset amplifies the curve recorded at $\mu_0 H = 0.9\,$T ($\mu_0 H_\text{int} \sim 0.7\,$T), confirming that SkL is a thermodynamic phase. Note the characteristic two-step ordering in $H = 0$.}
    \label{fig:fig2}
  \end{center}
\end{figure}
The recent discovery of topologically protected skyrmion spin textures in the centrosymmetric intermetallics Gd$_2$PdSi$_3$ \cite{Kurumaji2019}, Gd$_3$Ru$_4$Al$_{12}$\cite{Hirschberger2018}, and GdRu$_2$Si$_2$\cite{Khanh2020} with Heisenberg Gd$^{3+}$ magnetic moments has introduced a new paradigm: next to SkLs in (i) ferromagnetic thin plates with strong dipolar interactions (typical vortex diameter $\lambda_\text{mag}\sim 100-4000\,$nm)\cite{Bogdanov1989,Yu2012,Nayak2017} as well as (ii) bulk magnets or interfaces with broken inversion symmetry and spin-orbit driven Dzyaloshinskii-Moriya interactions ($\lambda_\text{mag}= 5-200\,$nm)\cite{Muehlbauer2009,Yu2010,Matsuno2016}, these compounds form a third and as-yet little explored category; namely, magnets with highly symmetric lattices and competing interactions, where ultra-small spin textures of lateral dimensions around $\lambda_\text{mag} = 1-3\,$nm can be realized. As $\lambda_\text{mag}$ approaches the size of a single crystallographic unit cell, the interplay of electronic structure and the giant emergent magnetic field $\sim 500\,$T arising from the non-coplanar magnetic order gives rise to large responses in (thermo-) electric transport \cite{Kurumaji2019,Hirschberger2018,Hirschberger2019,Gao2019}. In addition, the high symmetry of the underlying crystal lattices promises new phenomenology, such as enhanced control of helicity and vorticity degrees of freedom \cite{Leonov2015} and degeneracy of various types of magnetic order \cite{Okubo2012}.

Progress in materials search was mostly motivated by early theoretical works, especially numerical simulations on the two-dimensional triangular lattice which predicted SkL formation both in the case of frustrated exchange interactions (e.g., in insulators) \cite{Okubo2012,Leonov2015,Gao2017} and in the case of conduction electron mediated Ruderman-Kittel-Kasuya-Yosida (RKKY) couplings \cite{Hayami2014,Lin2016,Hayami2016,Hayami2016b}. Our target compound is centrosymmetric Gd$_2$PdSi$_3$, which crystallizes in an AlB$_2$-type hexagonal structure with triangular lattice layers of Gd$^{3+}$ magnetic moments \cite{Tang2011}. In between the triangular planes, honeycomb nets of Pd/Si sites harbor conducting electronic states. It was pointed out early on that below the ordering temperature $T_N = 21\,$K, the sequence of phase transitions as a function of externally applied magnetic field $\mathbf{H}$ is highly anisotropic \cite{Saha1999,Frontzek2010}. Such a behavior arises naturally as follows: In $H = 0$, the wavevector of the magnetic modulation $\mathbf{q}$ is aligned within the hexagonal basal plane \cite{Frontzek2009}. $\mathbf{H}\,\parallelsum\,c$ maintains the high symmetry of the hexagonal lattice and preserves the degeneracy of three equivalent directions of $\mathbf{q}$. This configuration therefore allows for non-coplanar multi-$\mathbf{q}$ ordering, but any component of $\mathbf{H}$ in the hexagonal basal plane breaks the symmetry and is naively expected to destabilize multi-$\mathbf{q}$ order \cite{Kurumaji2019}. For example, the SkL collapses when $\mathbf{H}$ is tilted about $40^\circ$ or more away from the $c$ axis\cite{Kurumaji2019}.

Here, we carefully discuss the magnetic phase diagram of Gd$_2$PdSi$_3$ over a broad range of $T$ and $\mathbf{H}$, and show that isotropic behavior with respect to the field direction is recovered in moderately high magnetic fields. Our work reveals that the ordering vector $\mathbf{q}$ can be depinned from its initial orientation along the $a^*$ direction. In the process, we introduce the planar Hall effect as a powerful tool to determine the preferred orientation of $\mathbf{q}$ in this material class, and to confirm the single-$\mathbf{q}$ nature of the field-induced phases for $\mathbf{H}$ in the basal plane. The present findings indicate that control of the ordering vector with electrical current may be possible in centrosymmetric magnets with very short-range spin textures.

\begin{figure*}[htb]
  \begin{center}
	%trim: [from left edge, from bottom , from right edge, ..]
		\includegraphics[trim=0cm 0.cm 0cm 0.cm, width=1\linewidth]{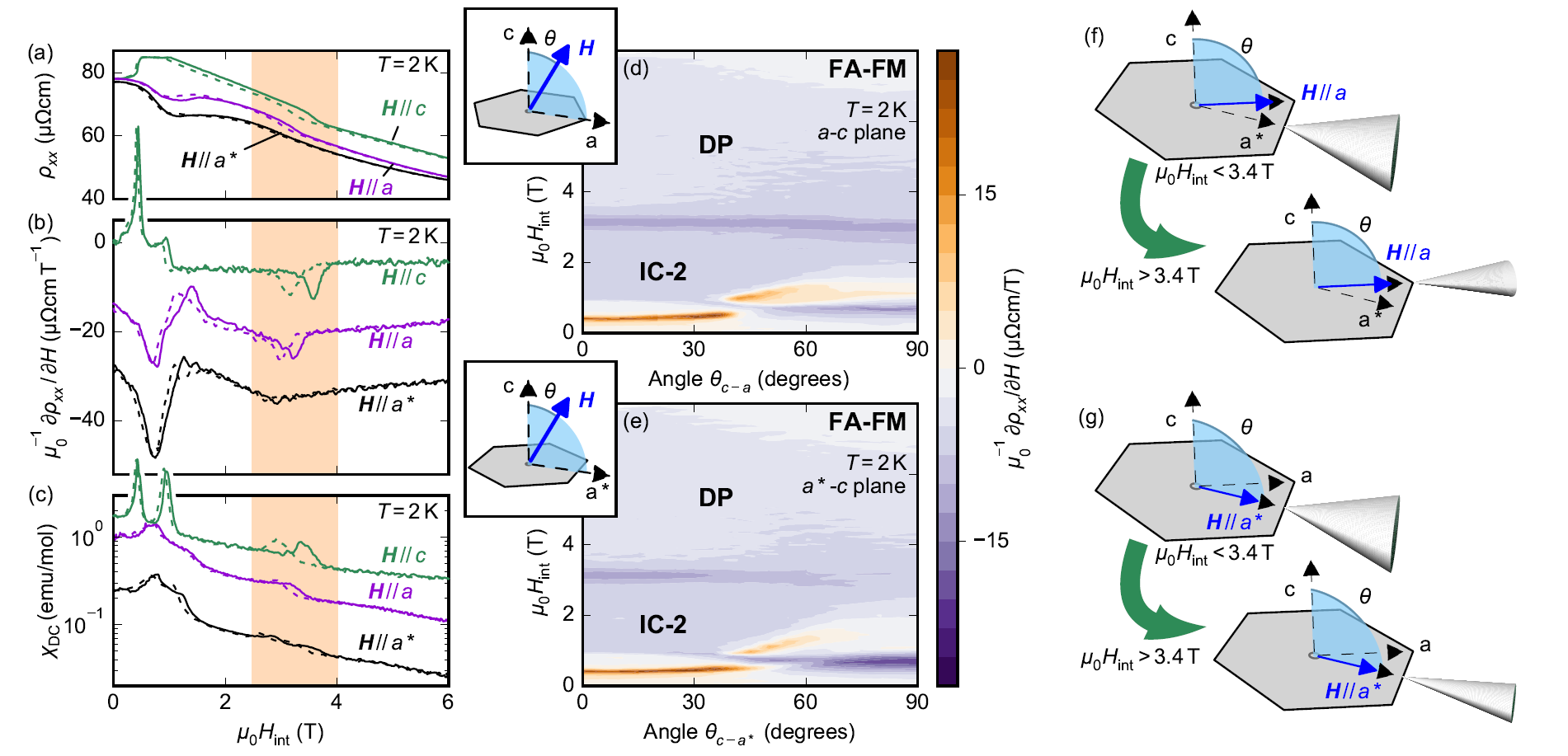}
    \caption[]{(color online). (a-c) Phase transitions for $\mathbf{H}$ along three high-symmetry directions, with the onset region of phase DP highlighted in red shading. Solid lines mark increasing, dashed lines decreasing $H$. (d, e) Phase diagram as constructed from the derivative of longitudinal resistivity $\mu_0^{-1}\,\partial \rho_{xx}/\partial H$. Insets depict the geometry of the respective rotation experiment. The dip in the signal around $\mu_0 H_\text{int} = 3.5 \,$T indicates the boundary between phases IC-2 and DP. The boundary remains sharp when rotating in the $c$-$a$ plane, but is washed out for $\mathbf{H}$ close to $a^*$. Panels (f, g) illustrate the proposed nature of the phase boundary for in-plane $\mathbf{H}$: $\mathbf{q}$ realigns only for $\mathbf{H}\,\parallelsum\,a$, leading to a sharp first-order transition in configuration (f). Blue shaded areas mark the rotation plane of $\mathbf{H}$, and the white cone describes the spin arrangement in the field-polarized spiral state.}
    \label{fig:fig3}
  \end{center}
\end{figure*}
We begin with the investigation of the magnetic phase diagram of Gd$_2$PdSi$_3$ for $\mathbf{H}$ along the $c$ direction (Fig. 1). The single crystals employed in this study were grown by the floating zone technique \cite{Kurumaji2019,Hirschberger2019} and cut and polished into the shape of a small cuboid (plate with current density $\mathbf{J}\,\parallelsum\,a$) for magnetization (resistivity) measurements. The effect of the demagnetization field was corrected according to $H_\text{int} = H - NM$, where $M$ is the magnetization density and $N$ is the demagnetization factor calculated in elliptical approximation (see Supplementary Information \cite{SI}). We extract the DC magnetic susceptibility as a field derivative viz. $\chi_\text{DC} = \partial M / \partial H$, where $M$ was measured in a commercial vibrating sample magnetometer. As compared to previous studies \cite{Kurumaji2019,Saha1999,Frontzek2010}, Figure 1 reports the existence of a new, as-yet unnoticed high-field phase (DP) and also clarifies that ordering in $H = 0$ proceeds through a sequence of two consecutive phase transitions [Fig. 1 (a)]. In panels (a,b, f, g, h), we have further labeled and illustrated phases IC-1 (helical, possibly multi-$\mathbf{q}$, see below), the skyrmion lattice SkL, and the fan-like state IC-2, all of which were previously resolved using resonant elastic x-ray scattering (REXS) at the Gd-L$_2$ edge \cite{Kurumaji2019}. These three phases were found to be incommensurate with the crystal lattice, with characteristic modulation length $\lambda_\text{mag} = 2.4\,$nm. The corresponding propagation vector in reciprocal lattice units is $\mathbf{q}_0 = (0.14, 0, 0)$, or linear combinations of $\mathbf{q}_0$ and symmetry-equivalents in phases IC-1 and SkL \cite{Kurumaji2019,Frontzek2009}. However, the REXS technique was unable to detect intensity at $\mathbf{q}_0$ in phase DP [Fig. 1 (c)], which is separated from IC-2 by a first-order phase transition (see hysteresis between up-and down field ramps in Fig. 1 (b) around $\mu_0 H_\text{int} = 3.5 \,$T). Note that at $4\,$T, the Zeeman energy of an individual Gd$^{3+}$ moment is roughly equal to $k_B T_N$, where $k_B$ is the Boltzmann constant. 

The phase boundaries, already apparent in $\chi_\text{DC}$, leave a somewhat sharper fingerprint in measurements of the electrical resistivity $\rho_{xx}$ [Fig. 1 (d)]. We cooled the sample slowly while keeping the magnitude of $H$ fixed; a sharp kink in $\rho_{xx}$, especially clear when $\mu_0 H > 0.9\,$T (corresponding to $\mu_0 H_\text{int} \sim 0.6\,$T, although $\mu_0 H_\text{int}$ changes with $T$), appears at the onset of long-range order in phases IC-2 or DP. Moreover, we have also confirmed the bulk nature of the phase transitions by specific-heat measurements $c_p(T)$ using a relaxation technique (Fig. 2). Note that $c_p$ was previously reported for a polycrystalline sample \cite{Mallik1998b}. The inset of this figure amplifies a curve which shows a kink upon entering the SkL phase. In $H = 0$, the onset of in-plane correlations, i.e. the sinusoidal state IC-2 (fan-like in higher fields), occurs before the development of full helical-type order in IC-1 [Fig. 1 (a)]. Such a two-step ordering process is also observed in the closely related material Gd$_3$Ru$_4$Al$_{12}$ \cite{Hirschberger2018}. 

Let $a$ and $a^*$ be vectors along the $[100]$ and $(100)$ directions of the hexagonal basal plane in real and reciprocal space, respectively. Further let us recall that these two directions are inequivalent from the viewpoint of symmetry. Transport properties and bulk magnetization at the lowest $T = 2\,$K for $\mathbf{H}$ along the $c$, $a$, and $a^*$ directions are reported side-by-side in Fig. 3 (a-c). From the raw data it is apparent that the sharp first-order transition at $\mu_0 H_\text{int} = 3.5-4\,$T is present only for $\mathbf{H}\,\parallelsum\, c$, $a$ (red shaded area in the plot). We expand on this point by showing the full data set for $\mu_0^{-1}\partial \rho_{xx}/\partial H$ as a function of field orientation and amplitude in Fig. 3 (d, e). The contour plot was assembled from isothermal ramps at $T = 2\,$K where the field magnitude $H$ was decreased continuously. The boundary between IC-2 and DP, marked by a minimum in $\mu_0^{-1}\partial \rho_{xx}/\partial H$, is conspicuously washed out in close vicinity of $\mathbf{H}\,\parallelsum\,a^*$. 

As noted above, phase IC-2 is known to be fan-like for $\mathbf{H}\,\parallelsum\,c$, i.e. there is an oscillatory (sinusoidal) in-plane magnetic moment $\mathbf{m}_{q}\perp \mathbf{q}$, $\mathbf{c}$ coexisting with uniform magnetization $\mathbf{M}\,\parallelsum\,\mathbf{H}$ [Fig. 1(g)]. When starting in IC-2 for $\mathbf{H}\,\parallelsum\,c$ and tilting the field towards the basal plane, no sharp phase transition could be detected; IC-2 for $\mathbf{H}\perp \mathbf{c}$ may therefore be assumed to be fan-like, or there could be a smooth cross-over towards a conical spiral as the field rotates. Figure 3 (f,g) illustrates the scenario proposed here for the transition between phases IC-2 and DP, assuming without loss of generality a conical spiral in phase IC-2 for $\mathbf{H}\perp \mathbf{c}$. A sharp first order transition occurs for $\mathbf{H}\,\parallelsum\,a$, when $\mathbf{q}$ suddenly changes its orientation to be parallel to $\mathbf{H}$. The transition is absent or very weak for $\mathbf{H}\,\parallelsum \,a^*$, because $\mathbf{H}$ is already parallel to $\mathbf{q}$ in this case. 

We further study changes of the magnetic order induced by in-plane fields via the planar Hall effect $\rho_{yx}^\text{PHE}$ (PHE), which measures the anisotropy between charge flow parallel and perpendicular to $\mathbf{q}$ in a modulated magnet via  the $H$-symmetric voltage drop perpendicular to $\mathbf{J}$ [Fig. 4 (a), upper right inset] \cite{Yokouchi2015}. Assuming that the resistivity tensor is diagonal in the Cartesian frame where $\mathbf{x}\,\parallelsum\,\mathbf{q}$ and $\mathbf{z} = c$, $\rho' = \text{diag}\left(\rho_\parallel, \rho_\perp\right)$ and through a simple coordinate transformation one derives $\rho_{yx}^\text{PHE} = \left(J/2\right)\cdot\left(\rho_\parallel-\rho_\perp\right)\sin\left(2\phi\right)$, where $\phi = \angle(\mathbf{q}, \mathbf{J})$. It is now easy to understand the distorted sinusoidal shape of $\rho_{yx}^\text{PHE}(\varphi)$ in Fig. 4 (a-d), where $\varphi = \angle\left(\mathbf{H}, \mathbf{J}\right)\neq \phi$: Due to pinning of $\mathbf{q}$ along $a^*$ and equivalent directions for $\mu_0 H_\text{int} < 3.5\,$T, the signal changes in abrupt steps [insets in Fig. 4 (a)]. Anisotropies $\left(\rho_\parallel-\rho_\perp\right)/\left(\rho_\parallel+\rho_\perp\right)\sim 2-5\,\%$ are easily resolved in our experiment.

\begin{figure}[t]
  \begin{center}
	%trim: [from left edge, from bottom , from right edge, ..]
		\includegraphics[trim=0.7cm 2.cm 0.7cm 1.cm, width=0.7\linewidth]{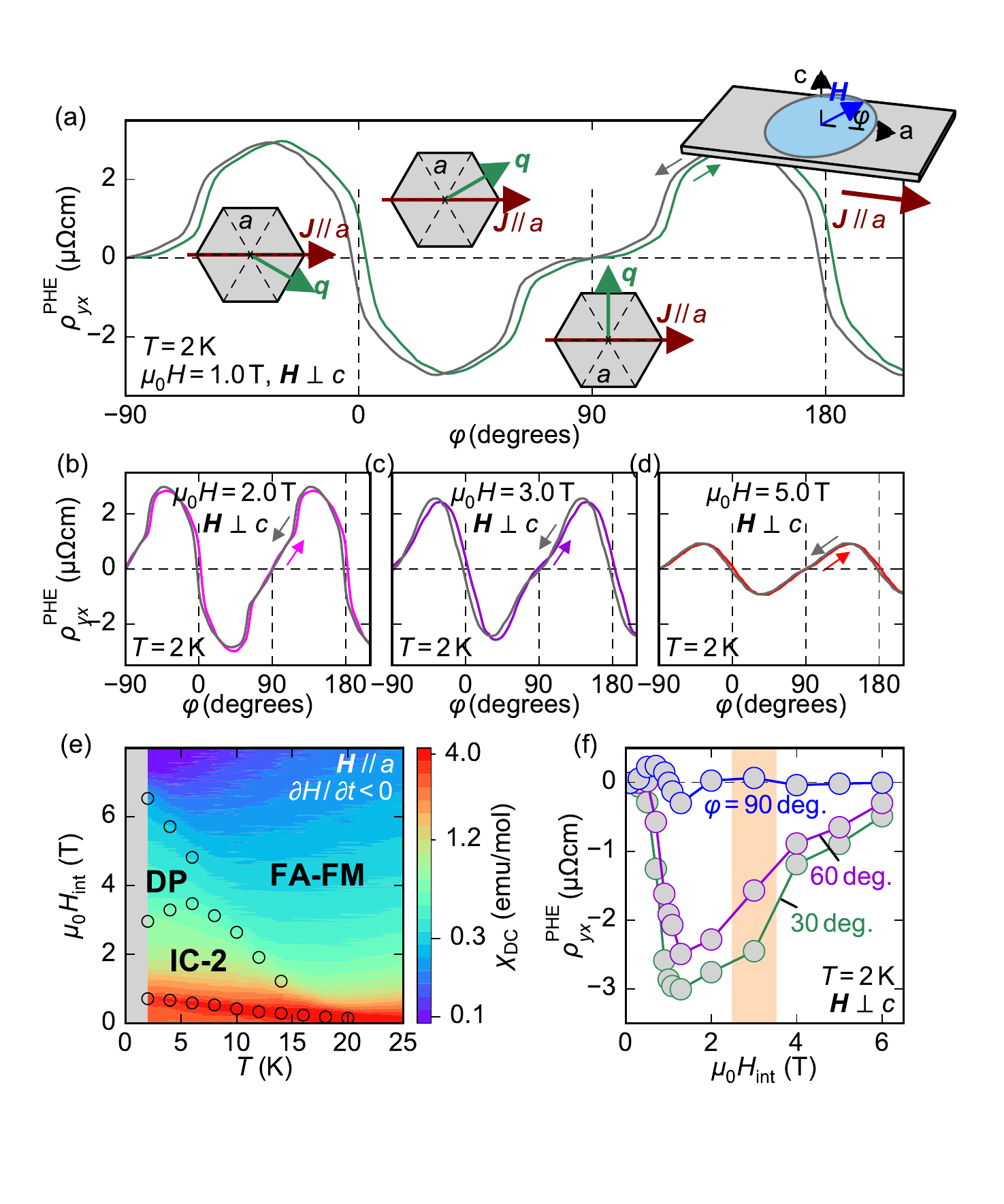}
    \caption[]{(color online). (a-d) Planar Hall effect after field cooling for selected values of $H$. Insets of (a): alignment of $\mathbf{q}$ and the electrical current density $\mathbf{J}$ at various $\varphi = \angle\left(\mathbf{H}, \mathbf{J}\right)$. Upper right inset: Geometry of this experiment, where $\mathbf{H}$ is rotated in the plane of the light blue disk, i.e. the hexagonal basal plane ($a$-$b$ plane). (e) Phase diagram for $\mathbf{H}\,\parallelsum\,a$ from magnetization measurements, analogous to Fig. 1(a). (f) $H$-dependence of the PHE amplitude at selected $\varphi$ values. The red shaded area indicates the regime of transition towards phase DP.}
    \label{fig:fig4}
  \end{center}
\end{figure} 

From the position of the flat plateau at $\varphi\sim 90^\circ$ (and not $\varphi\sim 0^\circ$), it is also apparent that the preferred orientation in IC-2 must be $\mathbf{q}\,\parallelsum\,a^*$. We demonstrate agreement of this result with REXS experiments in the Supplementary Information \cite{SI}. The PHE data demonstrates that in the modulated structure, current flows most easily perpendicular to $\mathbf{q}$. This is just as expected for the case of RKKY-mediated magnetic order, where a (partial) charge gap should open along $\mathbf{q}$. As $H$ increases, the step-like distortion of $\rho_{yx}^\text{PHE}(\varphi)$ is reduced, indicative of depinning of $\mathbf{q}$ from the $a^*$ axis [Fig. 4 (d)]. The maximum amplitude of the PHE changes sharply at the transition from IC-2 to DP, but no change of sign occurs [Fig. 4 (f)]. This shows that phase DP is still periodically modulated, with lower resistivity perpendicular to $\mathbf{q}$. To our knowledge, the PHE has never before been used as a sensor for the direction of $\mathbf{q}$ as well as its depinning transition.

We also observed\cite{SI} vanishing PHE in phase IC-1, even when preparing the sample in the field-cooled configuration with $\mathbf{H}$ in the hexagonal plane. Such behavior is in principle consistent with the hypothesis, put forward in Ref. \cite{Kurumaji2019}, that IC-1 is a multi-$\mathbf{q}$ state without net scalar spin chirality, i.e. a vortex-antivortex lattice [Fig. 1 (e)]. 

Isotropic phases in RKKY-type magnets are an active field of research where new classes of magnetic order, such as a multi-$\mathbf{q}$ ripple state on the honeycomb lattice, have found attention recently\cite{Shimokawa2019}. We compare two possible scenarios for phase DP in the present triangular lattice system: 

\textit{Commensurate ordering -} In the $R_2$PdSi$_3$ family, a crystallographic superstructure with $\mathbf{q}_C = (0.5, 0.5, 1/8)$ due to Pd/Si ordering on the honeycomb layer was discussed in previous work \cite{Tang2010}. Frontzek \etal{} reported locking of the magnetic order to the crystallographic distortion in for $R=$Ho, Tb, with $\mathbf{q}=\mathbf{q}_C$\cite{Frontzek2010b,Frontzek2010}. In contrast, our PHE experiments indicate that $\mathbf{q}\perp \mathbf{c}$ for $\mathbf{H}\perp \mathbf{c}$, inconsistent with $\mathbf{q}_C$-type ordering in phase DP. Supplementary neutron scattering experiments on $^{160}$Gd enriched single crystals also suggest that the $\mathbf{q}$-vector in the DP phase is still incommensurate\cite{SI}.

\textit{Tilted conical or fan-like order -} The modulation vector $\mathbf{q}$ likely takes an intermediate direction between the $a^*$ axis and $\mathbf{H}$ in phase DP. In the case where the magnetic field is in the hexagonal plane, the PHE data indicate continuous rotation of the ordering vector following the external field. In the case of $\mathbf{H}\,\parallelsum\,c$, neutron scattering is consistent with a gentle rotation of $\mathbf{q}$ away from the hexagonal plane in DP. However, the primary component of $\mathbf{q}$ is still along $a^*$ in this geometry\cite{SI}.

In the absence of strong single-ion anisotropy, the ground-state manifold in Gd$_2$PdSi$_3$ is determined by spin-spin couplings mediated through conducting electrons\cite{Nomoto2020}, and by dipolar energy. Note that the Fermi surface calculated for $R_2$PdSi$_3$ ($R=$Gd, Tb) has both cylindrical and more isotropic sheets, and strong pinning of $\mathbf{q}$ to the hexagonal plane was expected from previous numerical considerations\cite{Inosov2009,Nomoto2020}. In contrast, the presence of $\mathbf{q}\cdot c\neq 0$ in phase DP would indicate that exchange couplings between consecutive triangular lattice layers should be taken into account when modeling the magnetic order in Gd$_2$PdSi$_3$. The possibility of controlling $\mathbf{q}$ and the nanometric skyrmion lattice with electric current is left as a promising direction for future work.

\textit{Acknowledgments.} We thank T. Nomoto and S. Hayami for helpful discussions. Work at Photon Factory (KEK) and at J-PARC MLF was carried out under proposal numbers 2015S2-007 and 2019A0172, respectively. L.S. was funded by the German Academic Exchange Service (DAAD) via a PROMOS scholarship awarded by the German Federal Ministry of Education and Research (BMBF), with financial and administrative support by C. Pfleiderer. This work was partly supported by JST CREST Grant Number JPMJCR1874 (Japan). M. H. was supported as a Humboldt/JSPS International Research Fellow (18F18804).

\newpage
\section{Supplementary Material}
\beginsupplement

In the following, we discuss experimental procedures and present additional data in support of the conclusions in the main text. Further comments on crystal growth, annealing of single crystals, and sample preparation can be found in Ref. \cite{Kurumaji2019}.

%%%%%%%%%%%%%%%%%%%%%%%%%%%%%%%%%%%%%%%%%%%%%%%%%%%%%%%%%%%%%%%%%%%%%%
%%%%%%%%%%%%%%%%%%%%%%%%%%%%%%%%%
%%%%%%%%%%%%%%%%%%%%%%%%%%%%%%%%%
\section{Specific heat}
\label{sec:specheat}
%%%%%%%%%%%%%%%%%%%%%%%%%%%%%%%%%%%%%%%%%%

%%%%%%%%%%%%%%%%%%%%%%%%%%%%%%%%%%%%%%%%%%
%%%%%%%%%%%%%%%%%% FIG 6 %%%%%%%%%%%%%%%%%%%%%%
% trim: [from left edge, from bottom , from right edge, ..]
\begin{figure}[htb]
  \begin{center}
		\includegraphics[width=0.9\linewidth]{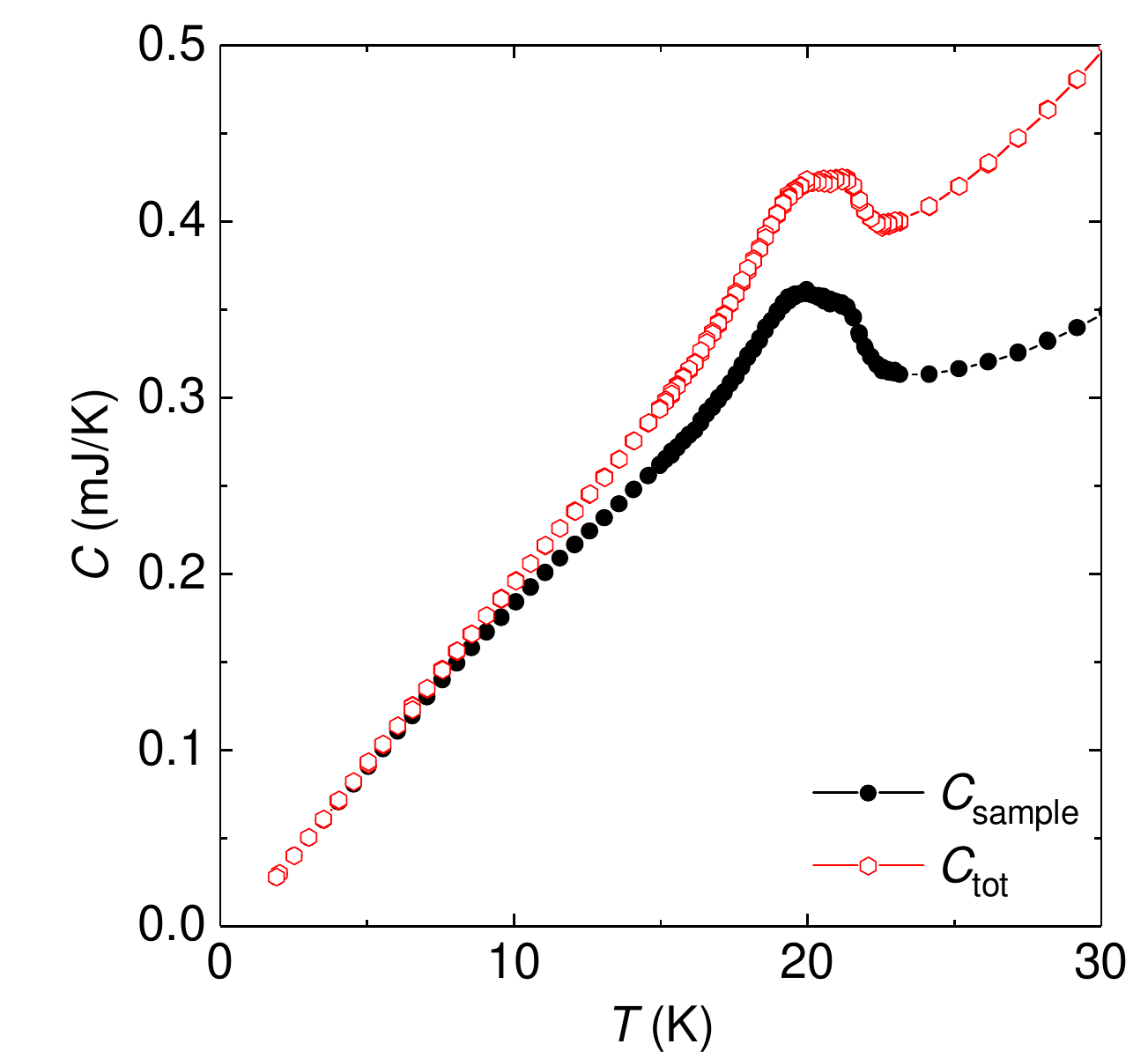}
    \caption[]{Comparison of sample heat capacity $C_\text{sample}$ and total heat capacity $C_\text{tot}$ (sample plus addenda) at low temperature $T$. Due to the high density of large magnetic moments and $T_N\sim20\,$K in this compound, the magnetic heat capacity of the sample dominates $C_\text{tot}$ in the temperature interval studied here.}
    \label{fig:hc_tot_data}
  \end{center}
\end{figure}
%%%%%%%%%%%%%%%%%%%%%%%%%%%%%%%%%%%%%%%%%%
%%%%%%%%%%%%%%%%%%%%%%%%%%%%%%%%%%%%%%%%%%%%%%%%%%%%%%%%%%%%%%%%%%%%%%

Measurements of specific heat $c_p$ were performed in a Quantum Design PPMS-9T cryostat, using the 'Heat Capacity' hardware option. For the $c_p$ measurements, the standard software package was employed. The $c_P$ experiment followed the standard routine, with $\sim2\,\%$ heat pulses applied to the sample. The heat capacity addenda $C_\text{addenda}$ were measured carefully as a function of temperature $T$. Note that $C_\text{addenda}$ is made up of contributions from both platform and grease.

Our measurements support the assumption that the magnetic torque is negligibly small for $\mathbf{H}\parallelsum\, c$ in this magnet with Heisenberg moments; there was no evidence for rotation or movement of the sample in high magnetic field.

\begin{figure}[htb]
  \begin{center}
		\includegraphics[width=0.9\linewidth]{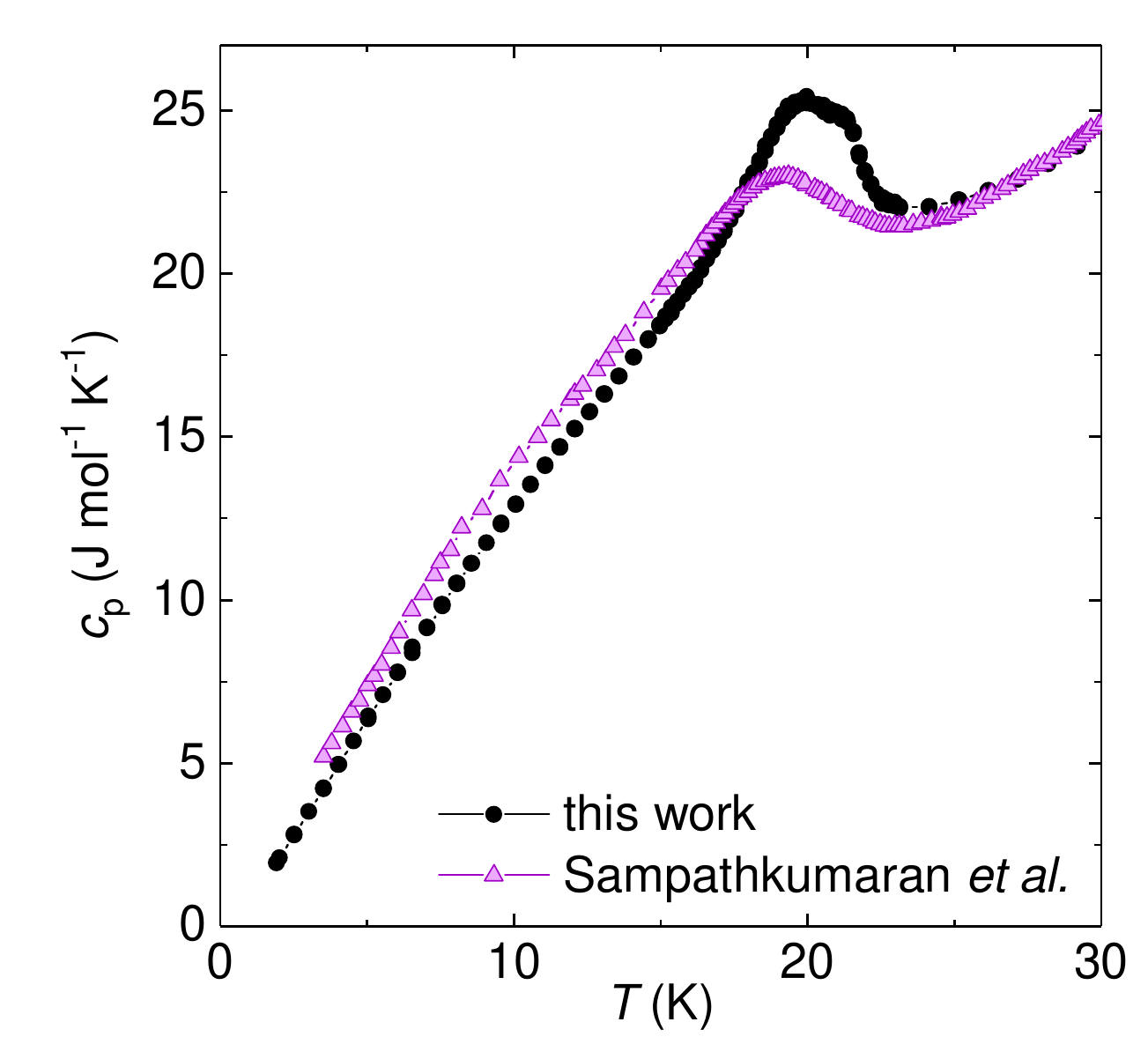}
    \caption[]{Comparison of specific heat $c_P(T)$ of Gd$_2$PdSi$_3$ for present work (single crystal, sample S1, black circles) and a previous report on a polycrystalline sample (manually extracted from Sampathkumaran \etal{} \cite{Sampathkumaran2000}, purple triangles). The external magnetic field was zero. Although the agreement is excellent in the paramagnetic phase, magnetic phase transitions are much sharper in the single crystal.}
    \label{fig:hc_compare_lit}
  \end{center}
\end{figure}

Fig. \ref{fig:hc_compare_lit} compares the results of our $c_p(T)$ measurements in $H=0$ to previous results obtained on a polycrystalline sample \cite{Sampathkumaran2000} (c.f. also \cite{Mallik1998b}). In the previous work, the phase transitions at $T_{N1}$ and $T_{N2}$ as described in the main text are washed out, forming a broad maximum in $c_P(T)$ below $20\,$K. However, we found quantitative agreement of the two data sets in the high- and low-$T$ regimes.

%%%%%%%%%%%%%%%%%% FIG 3 %%%%%%%%%%%%%%%%%%%%%%
% trim: [from left edge, from bottom , from right edge, ..]

%%%%%%%%%%%%%%%%%%%%%%%%%%%%%%%%%%%%%%%%%%
%%%%%%%%%%%%%%%%%%%%%%%%%%%%%%%%%%%%%%%%%%%%%%%%%%%%%%%%%%%%%%%%%%%%%%

%%%%%%%%%%%%%%%%%%%%%%%%%%%%%%%%%%%%%%%%%%%%%%%%%%%%%%%%%%%%%%%%%%%%%%
%%%%%%%%%%%%%%%%%%%%%%%%%%%%%%%%%
%%%%%%%%%%%%%%%%%%%%%%%%%%%%%%%%%
\section{Demagnetization correction in rotated magnetic field}
\label{sec:demag}
%%%%%%%%%%%%%%%%%%%%%%%%%%%%%%%%%%%%%%%%%%
%%%%%%%%%%%%%%%%%% FIG 4 %%%%%%%%%%%%%%%%%%%%%%
% trim: [from left edge, from bottom , from right edge, ..]
\begin{figure}[htb]
  \begin{center}
		\includegraphics[clip, trim=1.7cm 7.2cm 15.5cm 0.5cm, width=1.0\linewidth]{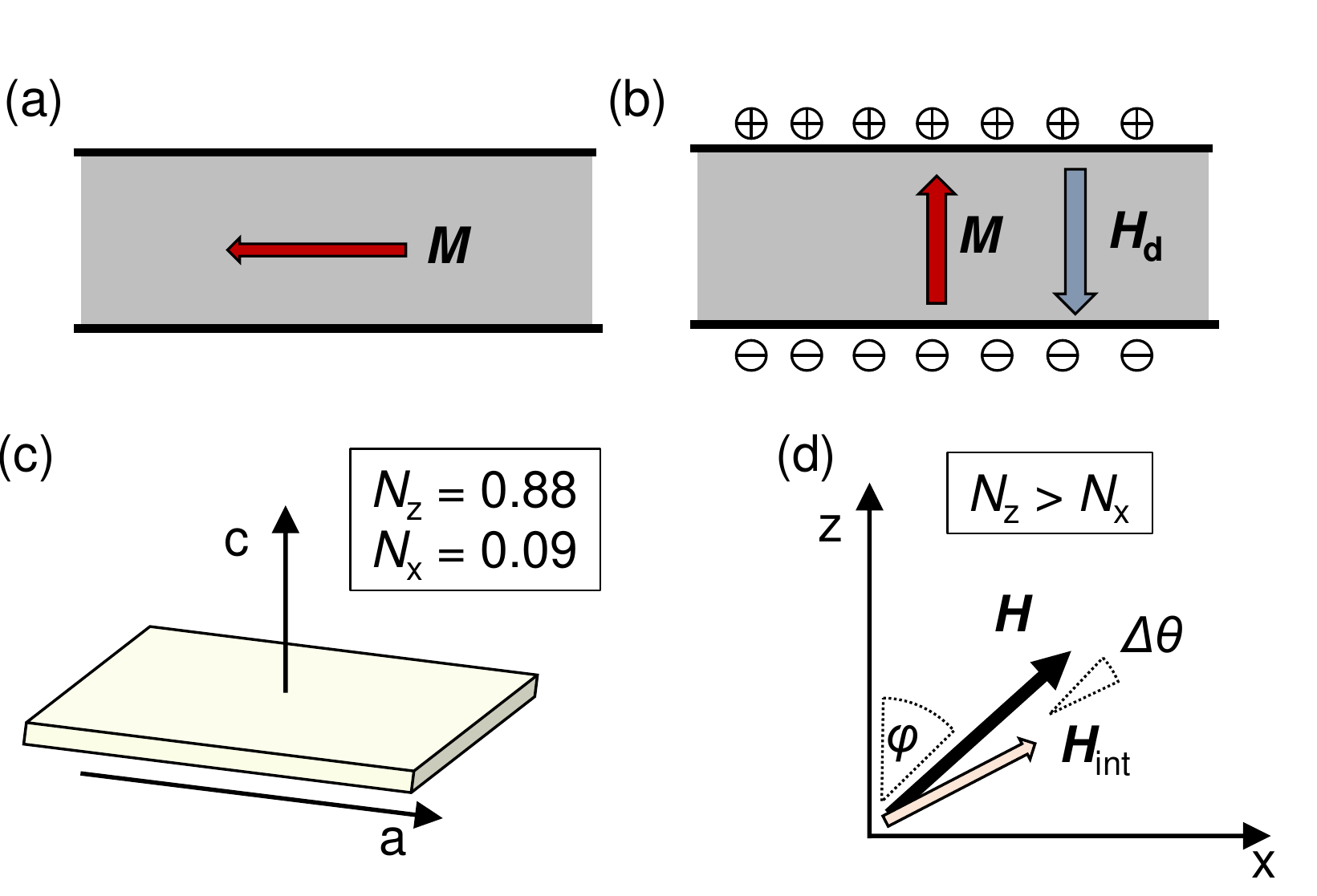}
    \caption[]{Demagnetization correction in thin plates of Gd$_2$PdSi$_3$. (a) In the case of an infinite plate, the demagnetization factor is zero for magnetization $\mathbf{M}$ aligned in the plane. (b) If $\mathbf{M}$ is perpendicular to the sample plane, magnetic charges at the interface between sample and vacuum (plus and minus signs in the figure) create a demagnetization field $\mathbf{H}_d$ which partially cancels the externally applied magnetic field $\mathbf{H}$. (c) Typical sample geometry used for transport experiments (sample S3). The crystallographic $c$-axis is perpendicular to the sample plane. (d) Let $\Delta \theta$ be the angle between internal magnetic field $\mathbf{H}_\text{int}$ and external magnetic field $\mathbf{H}$, and $\varphi$ be the angle between $\mathbf{H}$ and the $c$-axis. In the case where the demagnetizaton tensor is diagonal, and $N_c>N_a$, $\mathbf{H}$ is closer to the $c$-axis than $\mathbf{H}_\text{int}$.}
    \label{fig:demag_cor}
  \end{center}
\end{figure}
%%%%%%%%%%%%%%%%%%%%%%%%%%%%%%%%%%%%%%%%%%
%%%%%%%%%%%%%%%%%%%%%%%%%%%%%%%%%%%%%%%%%%%%%%%%%%%%%%%%%%%%%%%%%%%%%%
The origin of the demagnetizing field $\mathbf{H}_d$ is illustrated using the example of an infinite plate of homogeneous magnetization $\mathbf{M}$, embedded in vacuum with zero magnetization (c.f. Ref. \cite{Blundell2001}). Maxwell's equation for the magnetic field $\mathbf{B}=\mu_0\left(\mathbf{H}+\mathbf{M}\right)$ reads (SI units)
\begin{align}
\nabla \mathbf{B} &= 0\\
\nabla\mathbf{H} &= -\nabla\mathbf{M}
\end{align}
If the magnetization is completely parallel to the sample plane (Fig. \ref{fig:demag_cor} (a)), $\nabla\cdot \mathbf{M}= \nabla\cdot \mathbf{H}=0$ at the surface. If however $\mathbf{M}$ has a component perpendicular to the surface of the plate (Fig. \ref{fig:demag_cor} (b)), that component must change abruptly at the interface. This implies the presence of magnetic charges, i.e. sources of the magnetization field, shown as plus and minus signs in Fig. \ref{fig:demag_cor} (b). The demagnetization field $\mathbf{H}_d$ resulting from these charges tends to reduce the 'true' field viz. $\mathbf{H}_\text{int} = \mathbf{H} - \mathbf{H}_d$. 

It is common practice to approximate the shape of a cuboid or plate-like sample by an ellipsoid with principal axes corresponding to the sample's experimentally determined width, length, and thickness. In the case of an ellipsoid, the demagnetizing field is described by the linear tensor relation $\mathbf{H}_d = \bar{N}\cdot \mathbf{M}$, where the demagnetization tensor $\bar{N}$ is represented by a diagonal matrix and depends exclusively on the relative magnitude of the principal axes of the ellipsoid \cite{Osborn1945}. Note that the trace obeys $\text{tr}\left(\bar{N}\right) = 1$. The following paragraphs explain our approach towards the demagnetization correction in cases where the external magnetic field is not aligned with one of the principal axes of the ellipsoid. We make several approximations in the analysis of this generalized case.

Consider the geometry of Fig. \ref{fig:demag_cor} (d), where $\mathbf{H}$ is applied within a plane spanned by two principal directions $x$ and $z$, with $N_z>N_x$. This is the case, for example, in our transport measurements where $x=a^{*}$ and $z = c$ (c.f. Fig. 3, main text and Fig. \ref{fig:demag_cor} (c)). Assuming that approximately $\mathbf{M}\parallelsum\,\mathbf{H}_\text{int}$ in Gd$_2$PdSi$_3$, a material with Heisenberg moments, we write
\begin{equation}
\label{eq:demag}
\left(\begin{matrix}H\cos\varphi\\0\\H\sin\varphi\end{matrix}\right) = \left(\begin{matrix}\left(H_\text{int}+N_xM\right)\cos\theta\\0\\\left(H_\text{int} + N_zM\right)\sin\theta\end{matrix}\right)
\end{equation}
where $H_\text{int} = \left|\mathbf{H}_\text{int}\right|$ and $\theta = \varphi + \Delta \theta$ (Fig. \ref{fig:demag_cor}). In a further step, we use the scalar relation $M(H)$ measured for $\mathbf{H}\parallelsum\mathbf{M}\parallelsum a$, assuming an isotropic relationship between the applied field and the magnetization. This assumption entails an error of $<10\,\%$. Combining the $x$ and $z$ components of Eq. \ref{eq:demag}, we derive 
\begin{equation}
\theta = \tan^{-1}\left(\tan\varphi\,\frac{H_\text{int}+N_zM}{H_\text{int}+N_xM}\right)
\end{equation}
Using this, the relationship $H\left(H_\text{int}\right)$ at fixed $\varphi$ is now easily determined. As a single-valued function, it can be inverted to provide a mapping between external and internal magnetic field.

%%%%%%%%%%%%%%%%%%%%%%%%%%%%%%%%%%%%%%%%%%%%%%%%%%%%%%%%%%%%%%%%%%%%%%
\begin{table*}
	\centering
	\begin{tabular}{|c|c|c|c|c|c|c|c|}
	\hline
Sample \#	& measurement	& shape& 	batch \#	& $N_H$ &comments \\
	\hline\hline
S1	& Magnetization	& cuboid	& 1 &  $0.37$ & \\%S8B
S2	& Specific heat	& cuboid	& 1 &  $0.44$ & \\%S22 (?)
S3	& Resistivity measurements	& plate	& 1 &  $0.85$ & \\ %R122, S11, same as TN27-4
S4	& Planar Hall effect	& plate	& 1 &  $<0.1$ & \\%R80, planar Hall, S9
S5	& Resonant x-ray scattering	& thick plate	& 1 & $0.14$ & \\
S6	& Neutron scattering	& thick plates	& 2 &  $\sim 0.15$ & $^{160}$Gd enriched\\

	\hline

	\end{tabular}
	\caption{Samples (single crystals) used in this study. The crystal was grown by the optical floating zone technique in a high-vacuum compatible mirror furnace and subsequently annealed around $750^\circ\,$C and quenched in water \cite{Kurumaji2019}. $N_H$ is the demagnetization factor, calculated in elliptical approximation, for the principal axis parallel to the external field $H$. Batch 2 was enriched with $^{160}$Gd isotope to reduce the absorption cross-section in neutron scattering.}
  \label{tab:samples}
\end{table*}
%%%%%%%%%%%%%%%%%%%%%%%%%%%%%%%%%%%%%%%%%%%%%%%%%%%%%%%%%%%%%%%%%%%%%%

%%%%%%%%%%%%%%%%%%%%%%%%%%%%%%%%%%%%%%%%%%%%%%%%%%%%%%%%%%%%%%%%%%%%%%
%%%%%%%%%%%%%%%%%%%%%%%%%%%%%%%%%
%%%%%%%%%%%%%%%%%%%%%%%%%%%%%%%%%
\section{Hysteresis of magnetic phase diagram}
\label{sec:pd_hys}
%%%%%%%%%%%%%%%%%%%%%%%%%%%%%%%%%%%%%%%%%%
%%%%%%%%%%%%%%%%%% FIG X %%%%%%%%%%%%%%%%%%%%%%
% trim: [from left edge, from bottom , from right edge, ..]
\begin{figure}[htb]
  \begin{center}
		\includegraphics[width=1.\linewidth]{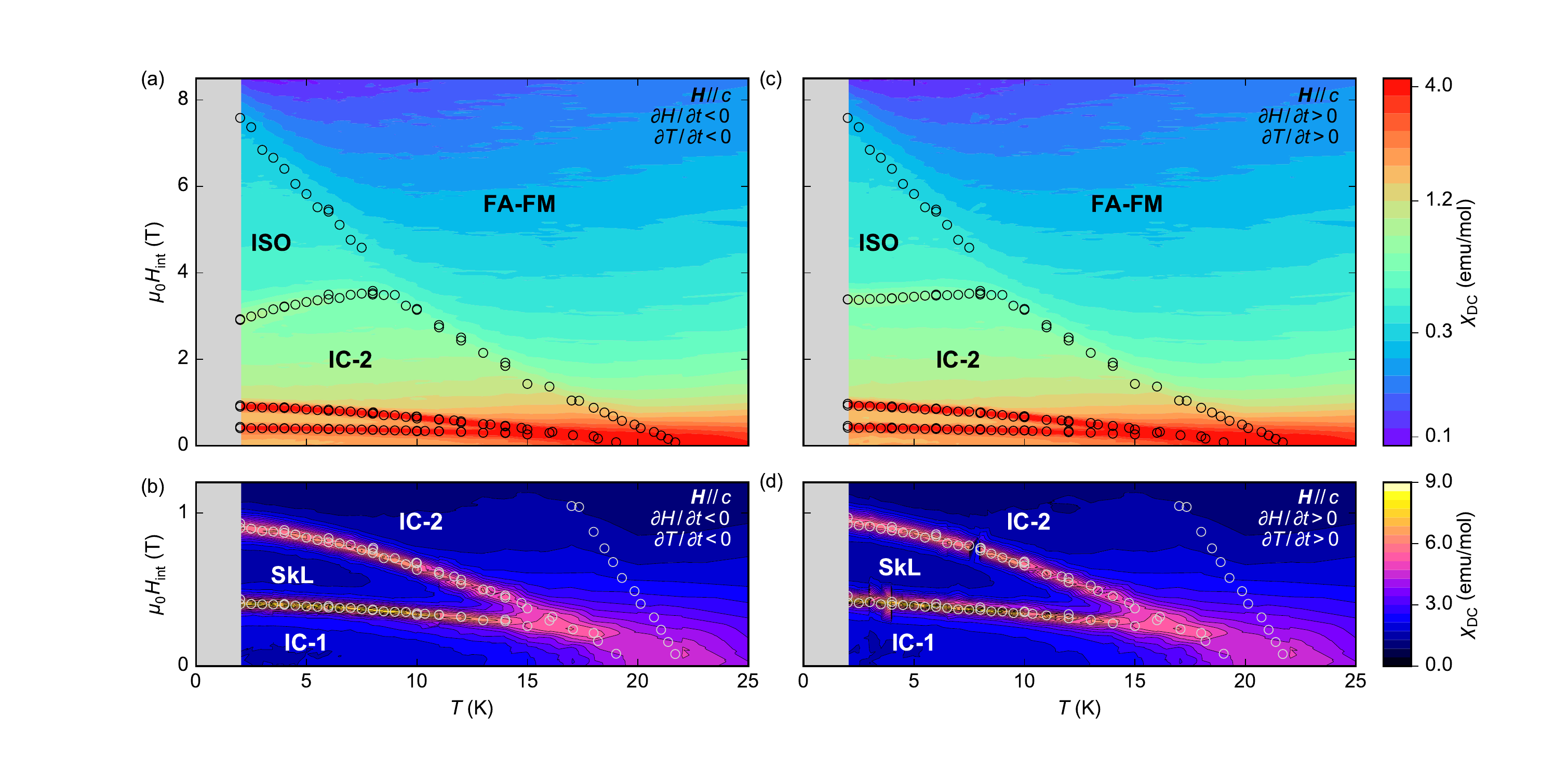}
    \caption[]{Comparison of magnetic phase diagram with magnetic field along the $c$-axis for (a, b) decreasing and (c, d) increasing magnetic field. Data in (a,b) is identical to Fig. 1 of the main text. Some hysteresis is observed at the phase transitions.}
    \label{fig:si_pd}
  \end{center}
\end{figure}
%%%%%%%%%%%%%%%%%%%%%%%%%%%%%%%%%%%%%%%%%
%%%%%%%%%%%%%%%%%%%%%%%%%%%%%%%%%%%%%%%%%%%%%%%%%%%%%%%%%%%%%%%%%%%%%%
Figure \ref{fig:si_pd} shows magnetic phase diagrams for Gd$_2$PdSi$_3$ comparing both increasing and decreasing magnetic field, for $\mathbf{H}\,\parallelsum\,c$. At the lowest temperatures, especially the transitions at $3-4\,$T show significant hysteresis. This is also apparent, e.g., from Fig. 1 (c) of the main text.

%%%%%%%%%%%%%%%%%%%%%%%%%%%%%%%%%%%%%%%%%%%%%%%%%%%%%%%%%%%%%%%%%%%%%%
%%%%%%%%%%%%%%%%%%%%%%%%%%%%%%%%%
%%%%%%%%%%%%%%%%%%%%%%%%%%%%%%%%%
\section{Supporting data of Planar Hall Effect}
\label{sec:si_phe}
%%%%%%%%%%%%%%%%%%%%%%%%%%%%%%%%%%%%%%%%%%
%%%%%%%%%%%%%%%%%% FIG X %%%%%%%%%%%%%%%%%%%%%%
% trim: [from left edge, from bottom , from right edge, ..]
\begin{figure}[htb]
  \begin{center}
		\includegraphics[width=1.\linewidth]{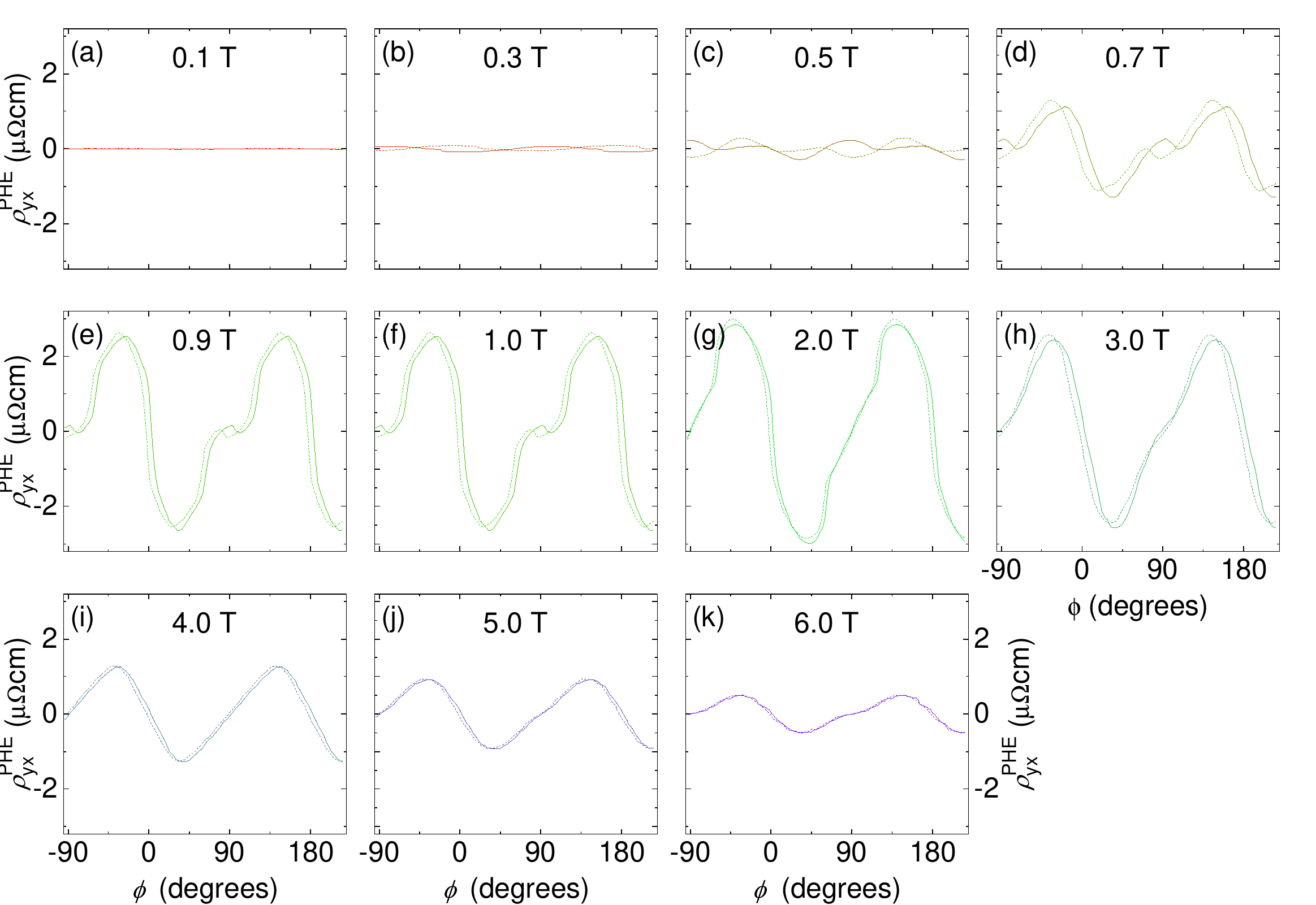}
    \caption[]{Extended data for planar Hall effect (PHE) measurements on Gd$_2$PdSi$_3$. External magnetic field $\mu_0 H$ is marked within each panel. Because the field is applied within the plane of the sample, the demagnetization correction is relatively small, $H \sim H_\text{int}$. The PHE $\rho_{yx}^\text{PHE}$ experiences a rapid onset around $0.5\,$T, the critical field for overcoming the IC-1 phase. The temperature of the sample was $T = 2\,$K.}
    \label{fig:si_phe}
  \end{center}
\end{figure}
%%%%%%%%%%%%%%%%%%%%%%%%%%%%%%%%%%%%%%%%%%
%%%%%%%%%%%%%%%%%%%%%%%%%%%%%%%%%%%%%%%%%%%%%%%%%%%%%%%%%%%%%%%%%%%%%%
In Fig. \ref{fig:si_phe}, we show extended data for planar Hall effect measurements in Gd$_2$PdSi$_3$, which illustrate a point made in the main text with regards to the nature of the IC-1 phase. Although the multi-$\mathbf{q}$ nature of this phase remains to be detected by a microscopic technique in future work, the present transport experiments suggest the absence of resistivity anisotropy.

A possible alternate scenario for phase DP has been brought to our attention. Namely, DP may be a tilted cone state, where the orientation of the cone changes without a shift in the propagation direction ($\mathbf{q}$). To counter this hypothesis, we note
\begin{enumerate}
\item For $\mathbf{H}\,\parallelsum\, c$, the neutron and REXS data confirm that $\mathbf{q}$ is depinned from the hexagonal plane (Fig. 1, main text, and Fig. S7). 
\item The transition between IC-2 and DP occurs at roughly the same critical field for all directions of $\mathbf{H}$, and there are no subdivisions within DP when rotating $\mathbf{H}$. This indicates a depinning transition for $\mathbf{q}$, independent of the direction of $\mathbf{H}$. 
\item Dipolar energy, which is important in the present case due to large Gd$^{3+}$ moments, does not favor a component of the oscillating moment parallel to $\mathbf{q}$. A tilt of the helical plane, while keeping $\mathbf{q}$ locked to the $a^{*}$ direction, would incur a significant energetic penalty due to the dipolar interaction. 
\item The planar Hall signal, taking a sinusoidal shape in DP, further supports the depinning scenario. If $\mathbf{q}$ were still pinned to the six $a^{*}$ directions of the hexagonal plane in phase DP, we expect anomalies in the PHE as $\mathbf{q}$ moves from one preferred direction to the next. Such anomalies of the PHE (jump- or step-like) are observed in IC-2, but not in DP (Fig. 4, main text).
\end{enumerate} 

We further note, although this was not demonstrated in the present study, that the PHE in the magnetically modulated phases (such as IC-1, SkL, IC-2, DP, etc) is expected to have opposite sign as compared to the case of the field-aligned ferromagnet. In the ferromagnetic case, we set $\mathbf{x}\,\parallelsum\,\mathbf{M}$ ($\mathbf{M}$ is the bulk magnetization) and again carry out a coordinate transform as in the main text. Then, the expression for the PHE is $\rho_{yx}^\text{PHE} = \left(J/2\right)\cdot\left(\rho_\parallel-\rho_\perp\right)\sin\left(2\phi\right)$, where now $\phi = \angle(\mathbf{M}, \mathbf{J})$. As both $\mathbf{M}$ and $\mathbf{q}$ (of a helical or fan-like state) follow $\mathbf{H}$ rather closely, the sign of the curve is determined by $\left(\rho_\parallel-\rho_\perp\right)$ in both cases. A distortion of the PHE curves, as discussed in the main text, occurs in the lower-field phases because $\mathbf{q}$ follows $\mathbf{H}$ only with some delay and in step-like jumps, due to pinning to preferred directions.

In the ferromagnetic case we typically have $\rho_\parallel<\rho_\perp$ in elevated magnetic fields, i.e. the longitudinal magnetoresistance is stronger as compared to the transverse one. This is in contrast to the $\mathbf{q}$-modulated phases, where $\rho_\parallel>\rho_\perp$ as a partial gap in the DOS is expected to be opened at $\varepsilon_F$ along the direction of $\mathbf{q}$. In fact, we have confirmed $\rho_\parallel>\rho_\perp$ in the ordered phases by measuring the angular dependence of the magnetoresistance (not shown). 

Local ion anisotropy of Gd$^{3+}$ magnetic spins is expected to generate some energetic difference between alignment of moments not only when comparing $c$ to the in-plane directions, but also within the hexagonal basal plane itself. Naively, this coupling between magnetic moments and the lattice may be seen as an impediment to the symmetry-based analysis of the PHE carried out in the main text. We justify our approach by emphasizing that local-ion anisotropy within the hexagonal basal plane is a higher order perturbative term, and is expected to be small as compared to exchange anisotropies and dipolar interactions.

%%%%%%%%%%%%%%%%%%%%%%%%%%%%%%%%%%%%%%%%%%%%%%%%%%%%%%%%%%%%%%%%%%%%%%
%%%%%%%%%%%%%%%%%%%%%%%%%%%%%%%%%
%%%%%%%%%%%%%%%%%%%%%%%%%%%%%%%%%
\section{Elastic neutron scattering on $^{160}$Gd enriched single crystals}
\label{sec:si_ens}

%%%%%%%%%%%%%%%%%%%%%%%%%%%%%%%%%%%%%%%%%%
%%%%%%%%%%%%%%%%%% FIG X %%%%%%%%%%%%%%%%%%%%%%
% trim: [from left edge, from bottom , from right edge, ..]
\begin{figure}[htb]
  \begin{center}
		\includegraphics[width=0.95\linewidth]{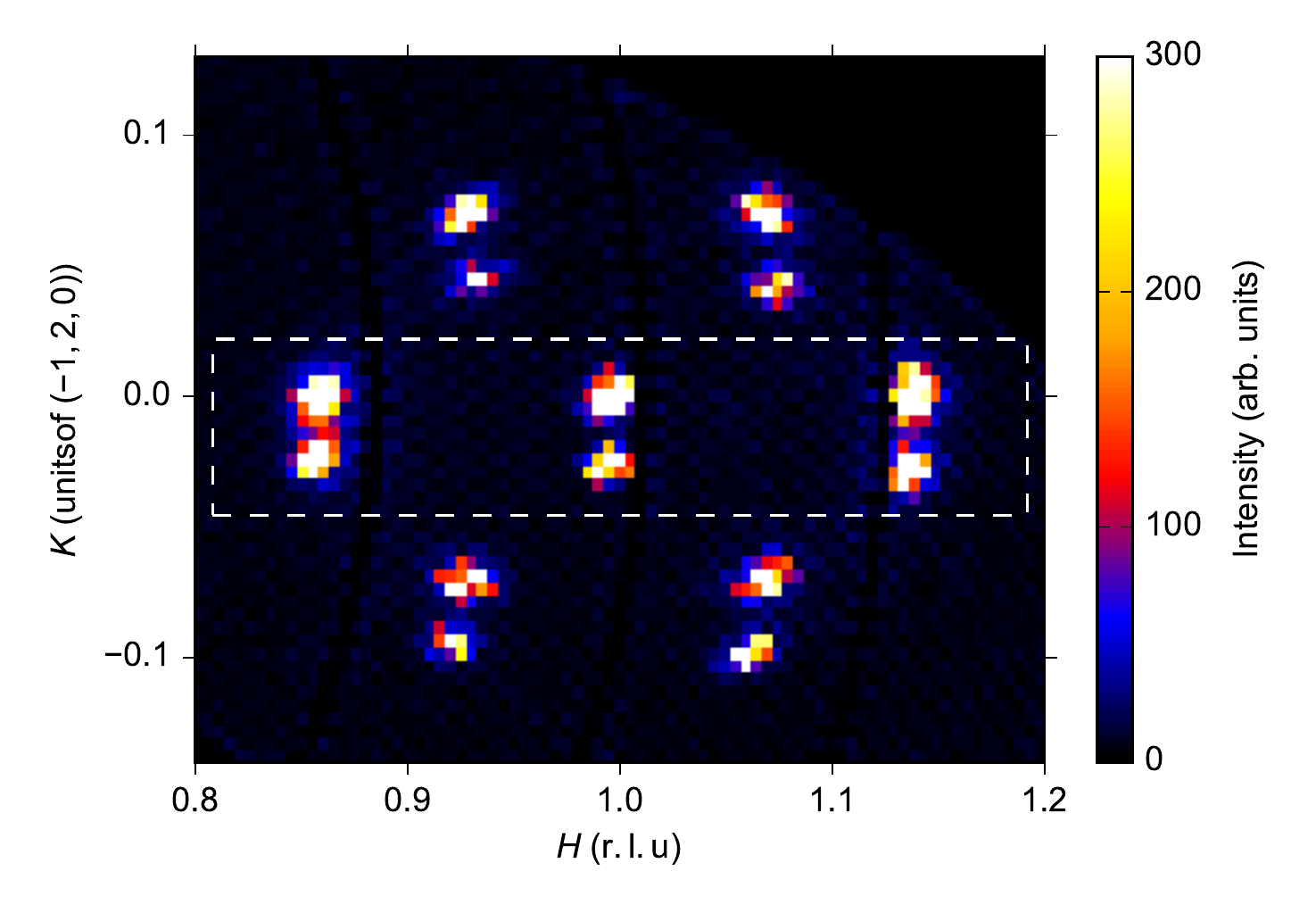}
    \caption[]{Elastic neutron scattering for $^{160}$Gd$_2$PdSi$_3$ at $H = 0$ and $ T = 3\,$K. Time-of-flight events with $\left|L\right|<0.02\,\mathrm{r.l.u.}$ and energy transfer $\left|E\right|<0.2\,$meV were binned to obtain this cut. The white dashed box indicates the $H$-$K$ area which was integrated to obtain the images in Fig. \ref{fig:si_neutron_lcut} (a-c). Several crystal pieces were coaligned for this experiment; this is apparent in the twinning effect of the magnetic pattern. See text for details.}
    \label{fig:si_neutron_hk0}
  \end{center}
\end{figure}
%%%%%%%%%%%%%%%%%%%%%%%%%%%%%%%%%%%%%%%%%%
%%%%%%%%%%%%%%%%%%%%%%%%%%%%%%%%%%%%%%%%%%%%%%%%%%%%%%%%%%%%%%%%%%%%%%

%%%%%%%%%%%%%%%%%%%%%%%%%%%%%%%%%%%%%%%%%%
%%%%%%%%%%%%%%%%%% FIG X %%%%%%%%%%%%%%%%%%%%%%
% trim: [from left edge, from bottom , from right edge, ..]
\begin{figure*}[htb]
  \begin{center}
		\includegraphics[width=0.95\linewidth]{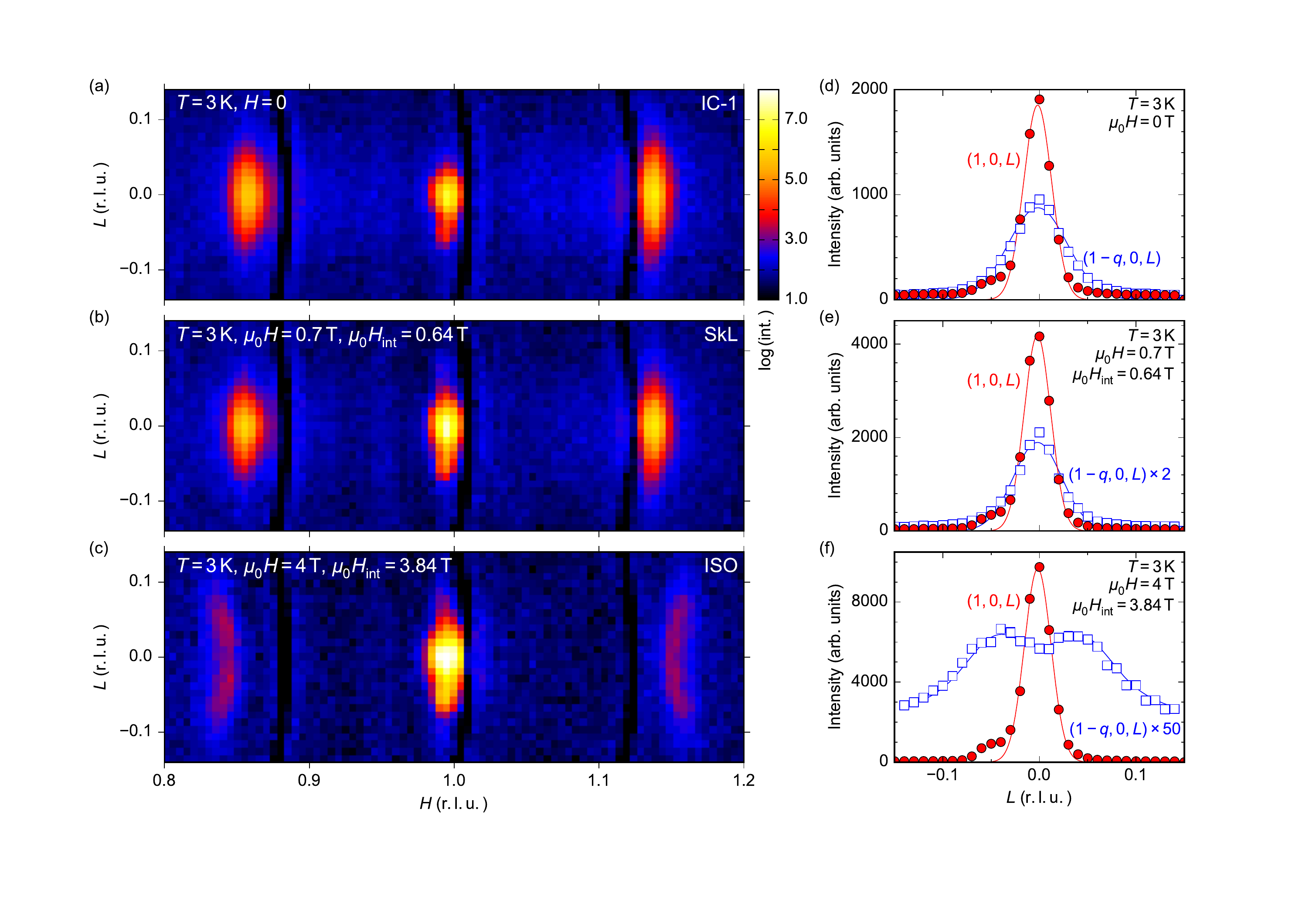}
    \caption[]{Elastic neutron scattering data for $^{160}$Gd$_2$PdSi$_3$, recorded with increasing magnetic field $\partial H / \partial t > 0$ along the crystallographic $c$-axis, at $T = 3\,$K. (a-c) Time-of-flight (TOF) data integrated over a slice in the $(H,K,L'=L)$ plane, where $L$ is a fixed value. Curves in panels (d-f) are line cuts of data from (a-c), fitted with Gaussian profiles (solid lines), or a double-Gaussian in the case of $(1-q, 0, L)$ in panel (f). See text for details.}
    \label{fig:si_neutron_lcut}
  \end{center}
\end{figure*}
%%%%%%%%%%%%%%%%%%%%%%%%%%%%%%%%%%%%%%%%%%
%%%%%%%%%%%%%%%%%%%%%%%%%%%%%%%%%%%%%%%%%%%%%%%%%%%%%%%%%%%%%%%%%%%%%%
We have grown large single crystals of $^{160}$Gd isotope-enriched Gd$_2$PdSi$_3$ using the floating zone technique. The samples were cut into plates of approximate dimensions $1\times5\times5\,$mm$^3$, with the largest face perpendicular to the $a^*$ direction. Several single crystalline pieces were coaligned using a Laue camera. The magnetic field was applied parallel to the $c$-axis, i.e. in the plane of the sample platelets. Therefore, the demagnetization factor is only $N\approx 0.15$ in elliptical approximation. 

Time of flight neutron scattering experiments were carried out at the cold-neutron disk chopper spectrometer AMATERAS~\cite{Nakajima2011} in the Materials and Life Science Experimental Facility (MLF) at the Japan Proton Accelerator Research Complex (J-PARC). The sample was loaded into a vertical-field $^4$He cryomagnet. We applied the magnetic field along the $c$-axis, and therefore the $(H,K,0)$ plane was chosen to be horizontal. A number of incident neutron energies $E_i$ were selected using the multi-$E_i$ method, and we here focus on the date measured with $E_i=3.60\,$meV.  The data was recorded at $T = 3\,$K, increasing the field in three steps starting from $H = 0$. The (quasi-) elastic scattering intensity within energy transfer range of $\left|E\right|<0.2\,$meV was extracted using the software UTSUSEMI \cite{Inamura2013}. 

Figure \ref{fig:si_neutron_hk0} shows intensities projected into the $(H,K,0)$ plane of reciprocal space. These data were obtained in zero magnetic field. Nuclear reflections are observed near the reciprocal lattice point $(1,0,0)$. Note the splitting of the reflection due to slight imperfection of the co-alignment of the crystals. Each nuclear reflection is accompanied by six-fold patterns of satellite magnetic reflections, which are indexed by the wavevector $(q,0,0)$ and its equivalents. Similar six-fold patterns are observed at all magnetic fields studied ($\mu_0 H = 0$, $0.7$, $4\,$T). Therefore, the magnetic ordering vector $\mathbf{q}$ is (nearly) aligned with the hexagonal basal plane even in phase DP. 

The elastic scattering data are integrated as a function of $K$, as marked by a white dashed rectangle in Fig. \ref{fig:si_neutron_hk0}. Through the integration, the signal from several domains (slightly misaligned sample pieces) can be added up. We display resulting profiles in the $(H, 0,L)$ plane for three magnetic phases in Fig. \ref{fig:si_neutron_lcut}. The intensity at $(1,0,0)$ increases continuously with $H$, as expected for the ferromagnetic component. Phase DP is characterized by a strong suppression of the magnetic signal amplitude at $\mathbf{q}$, together with an apparent split of the incommensurate magnetic reflection into two peaks along the $c^*$-axis (Fig. \ref{fig:si_neutron_lcut} (c)). 
%Moreover, there is no intensity at $(1, 0, 0.125)$ or $(0.5, 0.5, 0.125)$, in contrast to previous experiments on Ho$_2$PdSi$_3$ and Tb$_2$PdSi$_3$ \cite{Frontzek2004}\cite{Frontzek2010b}. 

We analyze the data by considering line-cuts of the $K$-integrated signal as a function of $L$. The nuclear / ferromagnetic intensity at $(1, 0, 0)$ is fitted with a Gaussian function. Its width and center position are unaffected by the magnetic field, indicating that no mechanical motion of the twinned sample occurred in our experiment. This is just as expected: Magnetic torques on our sample are expected to be very weak, due to the Heisenberg nature of the Gd$^{3+}$ moment. Furthermore, the magnetic intensity at the incommensurate location shrinks with $H$, and its width along $L$ increases significantly. The incommensurate reflections were fitted using a Gaussian profile in phases IC-1 and IC-2, while the profile in DP was fitted with two Gaussians on a Lorentzian background centered at $L = 0$. We found that the intensity profile along the $L$-direction is relatively broad at $\mu_0H=4\,$T. One possible interpretation is that the spin correlation length becomes shorter than those in the IC-1 and SkL phases. Another possible interpretation is that the amplitude of the $\mathbf{q}$-vector has a finite distribution in real space. To address this issue in more detail, further neutron scattering experiments will be necessary.

From the present data it is not possible to distinguish tilted fan and tilted conical orders in phase DP at $\mathbf{H}\,\parallelsum\,c$. No sharp phase transitions were observed as a function of rotation angle at $\mu_0H_\text{int} > 4\,$T. Nevertheless, a continuous cross-over in phase DP between depinned conical order in $\mathbf{H}\,\perp\,c$ and tilted fan-like order for $\mathbf{H}\,\parallelsum\,c$ is conceivable. Another possibility is that DP realizes a depinned / tilted conical state for all directions of $\mathbf{H}$. 

We further note that the six-fold scattering pattern in Fig. \ref{fig:si_neutron_hk0} may indicate either a multi-domain, single-$\mathbf{q}$ state, or a multi-$\mathbf{q}$ (potentially single domain) state. In the case of the DP phase, the bulk scattering data by itself is unable to distinguish these two cases. However, the phase diagrams in Fig. 3 of the main text clearly show that the critical field of the transition between IC-2 and DP is comparable for $\mathbf{H}\,\parallelsum\, c$, $a$, $a^{*}$. Moreover, within phase DP, we can rotate the field smoothly in all directions without meeting a phase boundary. It therefore appears somewhat unlikely that DP is a multi-$\mathbf{q}$ phase, even for $\mathbf{H}\,\parallelsum\,c$.

%%%%%%%%%%%%%%%%%%%%%%%%%%%%%%%%%
%%%%%%%%%%%%%%%%%%%%%%%%%%%%%%%%%
\section{REXS experiment for in-plane magnetic field}
\label{sec:si_rexs_ip}

%%%%%%%%%%%%%%%%%% FIG X %%%%%%%%%%%%%%%%%%%%%%
% trim: [from left edge, from bottom , from right edge, ..]
\begin{figure}[htb]
  \begin{center}
		\includegraphics[width=1.\linewidth]{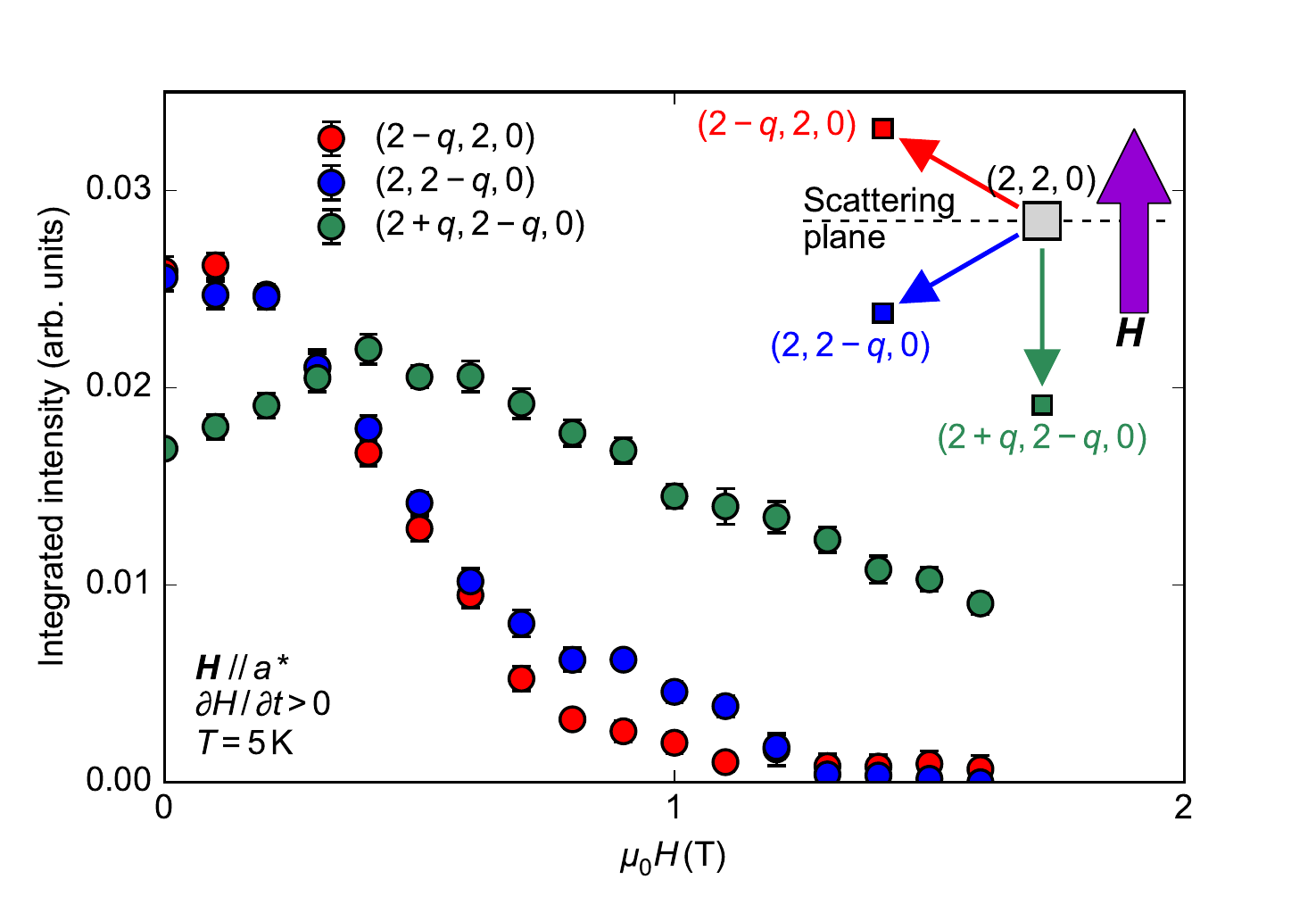}
    \caption[]{Resonant elastic x-ray scattering (REXS) at the Gd-L$_2$ edge on sample S5 with in-plane geometry ($\mathbf{H}\,\parallelsum\,a^*$). We show the magnetic intensity for three magnetic satellites around the $(2,2,0)$ Bragg reflection, as described in the inset. With increasing field, the intensity of two satellites is suppressed to zero, while the intensity of the satellite at $(2+q, 2-q, 0)$ remains finite in IC-2.}
    \label{fig:si_rexs_ip}
  \end{center}
\end{figure}
%%%%%%%%%%%%%%%%%%%%%%%%%%%%%%%%%%%%%%%%%%
In Fig. \ref{fig:si_rexs_ip}, we show supporting data which illustrates the single-$\mathbf{q}$ spiral (or fan) nature of the IC-2 phase when $\mathbf{H}\,\parallelsum\,a^*$, in good agreement with the interpretation of the planar Hall effect in the main text. Note the sketch in the inset of Fig. \ref{fig:si_rexs_ip}, illustrating the scattering geometry as well as the alignment of the scattering plane. The out-of-plane magnetic reflections where measured by $\chi$-scans under magnetic field along the $\left<1\bar{1}0\right>$ direction.

When increasing $\mathbf{H}$ along the in-plane direction, the intensity of two magnetic satellites decreases, while the satellite whose $\mathbf{q}$ is parallel to $\mathbf{H}$ begins to dominate the scattering signal. This is the expected behavior for a single-$\mathbf{q}$ state of fan-like or helical spiral nature. Note that in the fan-like or helical case, Zeeman energy prefers $\mathbf{q}\,\parallelsum\,\mathbf{H}$.


\begin{thebibliography}{99}
\bibitem{Kurumaji2019}T. Kurumaji, T. Nakajima, M. Hirschberger, A. Kikkawa, Y. Yamasaki, H. Sagayama, H. Nakao, Y. Taguchi, T. Arima, Y. Tokura, \textit{Skyrmion lattice with a giant topological Hall effect in a frustrated triangular-lattice magnet}. Science 10.1126/science.aau0968 (2019)
\bibitem{Hirschberger2018}M. Hirschberger, T. Nakajima, S. Gao, L. Peng, A. Kikkawa, T. Kurumaji, M. Kriener, Y. Yamasaki, H. Sagayama, H. Nakao, K. Ohishi, K. Kakurai, Y. Taguchi, X. Yu, T. Arima, Y.  Tokura, \textit{Skyrmion phase and competing magnetic orders on a breathing kagom{\'e} lattice}. Nature Communications \textbf{10}, 5831 (2019)
\bibitem{Khanh2020} N. D. Khanh, T. Nakajima, X. Z. Yu, S. Gao, K. Shibata, M. Hirschberger, Y. Yamasaki, H. Sagayama, H. Nakao, L. C. Peng, K. Nakajima, R. Takagi, T. Arima, Y. Tokura, S. Seki, \textit{Nanometric square skyrmion lattice in a centrosymmetric tetragonal magnet}, arXiv:2003.00626 (2020)
\bibitem{Bogdanov1989} A.N. Bogdanov and D.A. Yablonskii, \textit{Thermodynamically stable “vortices” in magnetically ordered crystals. The mixed state of magnets}, Journal of Experimental and Theoretical Physics \textbf{95}, 178 (1989)
\bibitem{Yu2012} X. Yu, M. Mostovoy, Y. Tokunaga, W. Zhang, K. Kimoto, Y. Matsui, Y. Kaneko, N. Nagaosa, and Y. Tokura, \textit{Magnetic stripes and skyrmions with helicity reversals}, Proceedings of the National Academy of Science \textbf{109} 8856-8860 (2012)
\bibitem{Nayak2017} A.K. Nayak, V. Kumar, T. Ma, P. Werner, E. Pippel, R. Sahoo, F. Damay, U.K. Rößler, C. Felser, and S.S.P. Parkin, \textit{Magnetic antiskyrmions above room temperature in tetragonal Heusler materials}, Nature \textbf{548}, 561–566 (2017)
\bibitem{Muehlbauer2009} S. M{\"u}hlbauer, B. Binz, F. Jonietz, C. Pfleiderer, A. Rosch, A. Neubauer, R. Georgii, P. B{\"o}ni, \textit{Skyrmion Lattice in a Chiral Magnet}. Science \textbf{323}, 915-919 (2009)
\bibitem{Yu2010} X.Z. Yu, Y. Onose, N. Kanazawa, J.H. Park, J.H. Han, Y. Matsui, N. Nagaosa, Y. Tokura. Real-space observation of a two-dimensional skyrmion crystal. Nature \textbf{456}, 901-904 (2010)
\bibitem{Matsuno2016}J. Matsuno, N. Ogawa, K. Yasuda, F. Kagawa, W. Koshibae, N. Nagaosa, Y. Tokura, M. Kawasaki, \textit{Interface-driven topological Hall effect in SrRuO$_3$-SrIrO$_3$ bilayer}. Science Advances \textbf{2}, e1600304 (2016)
\bibitem{Hirschberger2019} M. Hirschberger, L. Spitz, T. Nakajima, T. Kurumaji, A. Kikkawa, Y. Taguchi, and Y. Tokura, \textit{Topological Nernst effect of the two-dimensional skyrmion lattice.} arXiv:1910.06027 (2019)
\bibitem{Gao2019} S. Gao, M. Hirschberger, O. Zaharko, T. Nakajima, T. Kurumaji, A. Kikkawa, J. Shiogai, A. Tsukazaki, S. Kimura, S. Awaji, Y. Taguchi, T.-h. Arima, and Y. Tokura, \textit{Ordering phenomena of spin trimers accompanied by large geometrical Hall effect.} arXiv:1908.07728 (2019)
\bibitem{Leonov2015} A.O. Leonov and M. Mostovoy, \textit{Multiply periodic states and isolated skyrmions in an anisotropic frustrated magnet.} Nature Communications \textbf{6}, 8275 (2015)
\bibitem{Okubo2012} T. Okubo, S. Chung, and H. Kawamura, \textit{Multiple-q States and the Skyrmion Lattice of the Triangular-Lattice Heisenberg Antiferromagnet under Magnetic Fields.} Physical Review Letters \textbf{108}, 017206 (2012) 
\bibitem{Gao2017} S. Gao, O. Zaharko, V. Tsurkan, Y. Su, J.S. White, G.S. Tucker, B. Roessli, F. Bourdarot, R. Sibille, D. Chernyshov, T. Fennell, A. Loidl, and C. R{\"u}egg, \textit{Spiral spin-liquid and the emergence of a vortex-like state in MnSc$_2$S$_4$}. Nature Physics \textbf{13}, 157 (2017)
\bibitem{Hayami2014} S. Hayami and Y. Motome, \textit{Multiple-Q instability by (d-2)-dimensional connections of Fermi surfaces}, Physical Review B \textbf{90}, 060402(R) (2014)
\bibitem{Lin2016} S.-Z. Lin and S. Hayami, \textit{Ginzburg-Landau theory for skyrmions in inversion-symmetric magnets with competing interactions}. Physical Review B \textbf{93}, 064430 (2016)
\bibitem{Hayami2016} S. Hayami, S.-Z. Lin, and C.D. Batista, \textit{Bubble and skyrmion crystals in frustrated magnets with easy-axis anisotropy}. Physical Review B \textbf{93}, 184413 (2016)
\bibitem{Hayami2016b} S. Hayami, Y. Ozawa, and Y. Motome, \textit{Effective bilinear-biquadratic model for noncoplanar ordering in itinerant magnets}. Physical Review B \textbf{95}, 224424 (2017)
\bibitem{Tang2011} F. Tang, M. Frontzek, J. Dshemuchadse, T. Leisegang, M. Zschornak, R. Mietrach, J. U. Hoffmann, W. Loser, S. Gemming, D. C. Meyer, and M. Loewenhaupt, \textit{Crystallographic superstructure in $R_2$PdSi$_3$ compounds ($R$ = heavy rare earth)}, Phys. Rev. B \textbf{84}, 104105 (2011).
\bibitem{SI} Supplementary Information
\bibitem{Saha1999} S.R. Saha, H. Sugawara, T.D. Matsuda, H. Sato, R. Mallik, and E.V. Sampathkumaran, \textit{Magnetic anisotropy, first-order-like metamagnetic transitions, and large negative magnetoresistance in single-crystal Gd$_2$PdSi$_3$}, Physical Review B \textbf{60}, 12162 (1999)
\bibitem{Frontzek2010} M. Frontzek, F. Tang, P. Link, A. Schneidewind, J.M. Mignot, J.U. Hoffman, and M. Loewenhaupt, \textit{A Generic Phase Diagram for $R_2$PdSi$_3$ ($R =$ Heavy Rare Earth)?}, Journal of Physics: Conference Series, \textbf{251}, 012026 (2010)
\bibitem{Frontzek2009} M. Frontzek, \textit{Magnetic properties of $R_2$PdSi$_3$ ($R =$ heavy rare earth) compounds}, Doctoral thesis, Technische Universit{\"a}t Dresden (2009)
\bibitem{Mallik1998b} R. Mallik, E.V. Sampathkumaran, M. Strecker, and G. Wortmann, \textit{Observation of a minimum in the temperature-dependent electrical resistance above the magnetic-ordering temperature in Gd$_2$PdSi$_3$}, EPL (Europhysics Letters) \textbf{41}, 315 (1998)
\bibitem{Jensen1991} J. Jensen and A.R. Mackintosh, \textit{Rare Earth Magnetism}, Clarendon Press, Oxford (2001)
\bibitem{Yokouchi2015} T. Yokouchi, N. Kanazawa, A. Tsukazaki, Y. Kozuka, A. Kikkawa, Y. Taguchi, M. Kawasaki, M. Ichikawa, F. Kagawa, and Y. Tokura, \textit{Formation of In-plane Skyrmions in Epitaxial MnSi Thin Films as Revealed by Planar Hall Effect}, Journal of the Physical Society of Japan \textbf{84}, 104708 (2015) 
\bibitem{Pippard1989} A.B. Pippard, \textit{Magnetoresistance in Metals}, Cambridge University Press (1989)
\bibitem{Inosov2009}D.S. Inosov, D.V. Evtushinsky, A. Koitzsch, V.B. Zabolotnyy, S.V. Borisenko, A.A. Kordyuk, M. Frontzek, M. Loewenhaupt, W. L{\"o}ser, I. Mazilu, H. Bitterlich, G. Behr, J.-U. Hoffmann, R. Follath, B. B{\"u}chner, \textit{Electronic Structure and Nesting-Driven Enhancement of the RKKY Interaction at the Magnetic Ordering Propagation Vector in Gd$_2$PdSi$_3$ and Tb$_2$PdSi$_3$}. Physical Review Letters \textbf{102}, 046401 (2009)
\bibitem{Szytula1999} A. Szytula, M. Hofmann, B. Penc, M. Slaski, S. Majumdar, E.V. Sampathkumaran, A. Zygmunt, \textit{Magnetic behavior of $R_2$PdSi$_3$ compounds with $R=$Ce, Nd, Tb-Er}. Journal of Magnetism and Magnetic Materials \textbf{202}, 365-375 (1999)
\bibitem{Frontzek2004} M. Frontzek, A. Kreyssig, M. Doerr, J.-U. Hoffman, D. Hohlwein, H. Bitterlich, G. Behr, M. Loewenhaupt, \textit{Magnetic properties of Tb$_2$PdSi$_3$}. Physica B \textbf{350}, e187-e189 (2004)
\bibitem{Frontzek2010b} M. Frontzek, F. Tang, P. Link, A. Schneidewind, J.-U. Hoffman, J.-M. Mignot, and M. Loewenhaupt, \textit{Correlation between crystallographic superstructure and magnetic structures in finite magnetic fields: A neutron study on a single crystal of Ho$_2$PdSi$_3$}. Physical Review B \textbf{82}, 174401 (2010)
\bibitem{Mallik1998} R. Mallik, E.V. Sampathkumaran, M. Strecker, G. Wortmann, P.L. Paulose, and
Y. Ueda, \textit{Complex magnetism in a new alloy, Eu$_2$PdSi$_3$, with two crystallographically inequivalent sites}. Journal of Magnetism and Magnetic Materials \textbf{185}, L135-143 (1998) 
\bibitem{Tang2010} F. Tang, M. Frontzek, J. Dshemuchadse, T. Leisegang, M Zschornak, R. Mietrach, J.-U. Hoffmann, W. L{\"o}ser, S. Gemming, D.C. Meyer, and M. Loewenhaupt, \textit{Crystallographic superstructure in $R_2$PdSi$_3$ compounds ($R =$ heavy rare earth)}. Physical Review B \textbf{84}, 104105 (2011)
\bibitem{Shimokawa2019} T. Shimokawa and H. Kawamura, \textit{Ripple State in the Frustrated Honeycomb-Lattice Antiferromagnet}, Physical Review Letters \textbf{123}, 057202 (2019)
\bibitem{Nomoto2020} T. Nomoto, T. Koretsune, R. Arita, \textit{Formation mechanism of helical Q structure in Gd-based skyrmion materials}, arXiv:2003.13167 (2020)



%
%
%\bibitem{Nagaosa2013}N. Nagaosa, Y. Tokura, \textit{Topological properties and dynamics of magnetic skyrmions}. Nature Nanotechnology \textbf{8}, 899-911 (2013)
%\bibitem{Neubauer2009}A. Neubauer, C. Pfleiderer, B. Binz, A. Rosch, R. Ritz, P. Niklowitz, P. B{\"o}ni, Topological Hall Effect in the A Phase of MnSi. Physical Review Letters \textbf{102}, 186602 (2009)
%\bibitem{Lee2009}M. Lee, W. Kang, Y. Onose, Y. Tokura, N. P. Ong, \textit{Unusual Hall Effect Anomaly in MnSi under Pressure}. Physical Review Letters \textbf{102}, 186601 (2009)
%\bibitem{Ritz2013}R. Ritz, M. Halder, C. Franz, A. Bauer, M. Wagner, R. Bamler, A. Rosch, C. Pfleiderer, \textit{Giant generic topological Hall resistivity of MnSi under pressure}. Physical Review B \textbf{87}, 134424 (2013)
%\bibitem{Nagaosa2010}N. Nagaosa, J. Sinova, S. Onoda, A.H. MacDonald, N.P. Ong, \textit{Anomalous Hall effect}. Reviews of Modern Physics \textbf{82}, 1539 (2010)
%%\bibitem{Franz2014}C. Franz, F. Freimuth, A. Bauer, R. Ritz, C. Schnarr, C. Duvinage, T. Adams, S. Blügel, A. Rosch, Y. Mokrousov, C. Pfleiderer, \textit{Real-Space and Reciprocal-Space Berry Phases in the Hall Effect of Mn$_{1-x}$Fe$_x$Si}. Physical Review Letters \textbf{112}, 186601 (2014)
%%\bibitem{Oike2016}H. Oike, A. Kikkawa, N. Kanazawa, Y. Taguchi, M. Kawasaki, Y. Tokura, F. Kagawa, \textit{Interplay between topological and thermodynamic stability in a metastable magnetic skyrmion lattice}. Nature Physics \textbf{12}, 62–66 (2016)
%%\bibitem{Wang2018}L. Wang, Q. Feng, Y. Kim, R. Kim, K.H. Lee, S.D. Pollard, Y.J. Shin, H. Zhou, W. Peng, D. Lee, W. Meng, H. Yang, J.H. Han, M. Kim, Q. Lu, T.W. Noh, \textit{Ferroelectrically tunable magnetic skyrmions in ultrathin oxide heterostructures}. Nature Materials \textbf{17}, 1087 (2018)
%%\bibitem{Vistoli2019}L. Vistoli, W. Wang, A. Sander, Q. Zhu, B. Casals, R. Cichelero, A. Bart{\'e}l{\'e}my, S. Fusil, G. Herranz, S. Valencia, R. Abrudan, E. Weschke, K. Nakazawa, H. Kohno, J. Santamaria, W. Wu, V. Garcia, M. Bibes, \textit{Giant topological Hall effect in correlated oxide thin films}. Nature Physics \textbf{15}, 67–72 (2019)
%\bibitem{Ziman1979}J.M. Ziman, \textit{Principles of the Theory of Solids}. Second Edition, Cambridge University Press (1979)
%\bibitem{Smrcka1977}L. Smr\v{c}ka, P. Streda, \textit{Transport coefficients in strong magnetic fields}. Journal of Physics C: Solid State Physics \textbf{10}, 2153-2161 (1977)
%\bibitem{Tokunaga2015}Y. Tokunaga, X.Z. Yu, J.S. White, H.M. Rønnow, D. Morikawa, Y. Taguchi, Y. Tokura, \textit{A new class of chiral materials hosting magnetic skyrmions beyond room temperature}. Nature Communications \textbf{6}, 7638 (2015)
%\bibitem{Hanasaki2008}N. Hanasaki, K. Sano, Y. Onose, T. Ohtsuka, S. Iguchi, I. K{\'e}zsm{\'a}rki, S. Miyasaka, S. Onoda, N. Nagaosa, Y. Tokura, \textit{Anomalous Nernst Effects in Pyrochlore Molybdates with Spin Chirality}. Physical Review Letters \textbf{100}, 106601 (2008)
%\bibitem{Kanazawa2011}N. Kanazawa, Y. Onose, T. Arima, D. Okuyama, K. Ohoyama, S. Wakimoto, K. Kakurai, S. Ishiwata, Y. Tokura, \textit{Large Topological Hall Effect in a Short-Period Helimagnet MnGe}. Physical Review Letters \textbf{106}, 156603 (2011)
%\bibitem{Shiomi2013}Y. Shiomi, N. Kanazawa, K. Shibata, Y. Onose, Y. Tokura, Topological Nernst effect in a three-dimensional skyrmion-lattice phase. Physical Review B \textbf{88}, 064409 (2013)
%\bibitem{Schlitz2019}R. Schlitz, P. Swekis, A. Markou, H. Reichlova, M. Lammel, J. Gayles, A. Thomas, K. Nielsch, C. Felser, S.T.B. Goennenwein, \textit{All Electrical Access to Topological Transport Features in Mn$_{1.8}$PtSn Films}. Nano Letters \textbf{19}, 2366-2370 (2019)
%\bibitem{SI} Supplementary Information.
%
%

%\bibitem{Sakai2018}A. Sakai, Y.P. Mizuta, A.A. Nugroho, R. Sihombing, T. Koretsune, M.-T. Suzuki, N. Takemori, R. Ishii, D. Nishio-Hamane, R. Arita, P. Goswami, S. Nakatsuji, \textit{Giant anomalous Nernst effect and quantum-critical scaling in a ferromagnetic semimetal}, Nature Physics \textbf{14}, 1119–1124 (2018)
%\bibitem{Miyasato2007}T. Miyasato, N. Abe, T. Fujii, A. Asamitsu, S. Onoda, Y. Onose, N. Nagaosa, Y. Tokura, \textit{Crossover Behavior of the Anomalous Hall Effect and Anomalous Nernst Effect in Itinerant Ferromagnets}. Physical Review Letters \textbf{99}, 086602 (2007)
%
\end{thebibliography}

\begin{thebibliography}{99}
%\bibitem{Lovesey1996} S. W. Lovesey and S.P. Collins, \textit{X-ray Scattering and Absorption by Magnetic Materials}, Clarendon Press (1996)
\bibitem{Kurumaji2019}T. Kurumaji, T. Nakajima, M. Hirschberger, A. Kikkawa, Y. Yamasaki, H. Sagayama, H. Nakao, Y. Taguchi, T. Arima, Y. Tokura, \textit{Skyrmion lattice with a giant topological Hall effect in a frustrated triangular-lattice magnet}. Science 10.1126/science.aau0968 (2019)
\bibitem{Sampathkumaran2000} E. V. Sampathkumaran \etal{}, \textit{Magnetocaloric effect in Gd$_2$PdSi$_3$}, Applied Physics Letters \textbf{77}, 418 (2000)
\bibitem{Mallik1998b} R. Mallik, E.V. Sampathkumaran, M. Strecker, and G. Wortmann, \textit{Observation of a minimum in the temperature-dependent electrical resistance above the magnetic-ordering temperature in Gd$_2$PdSi$_3$}, EPL (Europhysics Letters) \textbf{41}, 315 (1998)
%\bibitem{Hirschberger2018} M. Hirschberger \etal, \textit{Skyrmion phase and competing magnetic orders on a breathing kagom{\'e} lattice}, arXiv:1812.02553 (2018)
%\bibitem{HirschbergerPhD} M. Hirschberger, \textit{Quasiparticle Excitations with Berry Curvature in Insulating Magnets and Weyl Semimetals}, PhD Thesis, Princeton University (2017)
\bibitem{Osborn1945} J. A. Osborn, \textit{Demagnetizing Factors of the General Ellipsoid}, Phys. Rev. \textbf{67}, 351 (1945)
\bibitem{Blundell2001} S. Blundell, \textit{Magnetism in Condensed Matter}, Oxford University Press (2001)
%\bibitem{Scheie2018} A. Scheie, J. Low Temp. Phys. \textbf{193}, 60 (2018)
\bibitem{Frontzek2004} M. Frontzek, A. Kreyssig, M. Doerr, J.-U. Hoffman, D. Hohlwein, H. Bitterlich, G. Behr, M. Loewenhaupt, \textit{Magnetic properties of Tb$_2$PdSi$_3$}. Physica B \textbf{350}, e187-e189 (2004)
\bibitem{Frontzek2010b} M. Frontzek, F. Tang, P. Link, A. Schneidewind, J.-U. Hoffman, J.-M. Mignot, and M. Loewenhaupt, \textit{Correlation between crystallographic superstructure and magnetic structures in finite magnetic fields: A neutron study on a single crystal of Ho$_2$PdSi$_3$}. Physical Review B \textbf{82}, 174401 (2010)
\bibitem{Inamura2013} Y. Inamura, T. Nakatani, J. Suzuki, and T. Otomo, \textit{Development status of software “Utsusemi” for chopper spectrometers at MLF, J-PARC}. Journal of the Physical Society of Japan \textbf{82}, SA031 (2013).
\bibitem{Nakajima2011} K. Nakajima, S. Ohira-Kawamura, T. Kikuchi, M. Nakamura, R. Kajimoto, Y. Inamura, N. Takahashi, K. Aizawa, K. Suzuya, K. Shibata, T. Nakatani, K. Soyama, R. Maruyama, H. Tanaka, W. Kambara, T. Iwahashi, Y. Itoh, T. Osakabe, S. Wakimoto, K. Kakurai, F. Maekawa, M. Harada, K. Oikawa, R.E. Lechner, F. Mezei, and M. Arai, \textit{AMATERAS: A Cold-Neutron Disk Chopper Spectrometer}. Proceedings of the International Workshop on Neutron Applications on Strongly Correlated Electron Systems 2011 (NASCES11), Journal of the Physical Society of Japan \textbf{80}, SB028 (2011) 
\end{thebibliography}
\end{document}